%% file: Fleeting_Prices_BN_v1.5.tex
\documentclass[graybox]{svmult}
\usepackage{amssymb}
\usepackage{amsmath}
\usepackage{multirow}

\input{preamble}
\input{tcilatex}
\begin{document}

\input{preintro}

\section{Introduction}

In recent years, Ole E. Barndorff-Nielsen has been working on a class of
stochastic models called integer-valued trawl processes. References include
\cite{BarndorffNielsen(11)}, \cite{BarndorffNielsenBenthVeraart(12)} and
\cite{BarndorffNielsenLundeShephardVeraart(14)}. These are flexible models
whose core randomness is driven by Poisson random measures. Trawl processes
are related to the up-stairs processes of \cite{WolpertTaqqu(05)} and the
random measure processes of \cite{WolpertBrown(11)}. Both of these processes
are stationary. \cite{BarndorffNielsenLundeShephardVeraart(14)} also brings
out the relationship between their processes and $\mathrm{M}/G/\infty $
queues (e.g. \cite{Lindley(56)}, \cite{Reynolds(68)} and \cite[Ch. 6.31]%
{Bartlett(78)}) and mixed moving average processes (e.g. \cite%
{SurgailisRosinskiMandrekarCambanis(93)}). Related discrete time count
models include \cite{CameronTrivedi(98)}, \cite{KedemFokianos(02)}, \cite%
{CuiLund(09)}, \cite{DavisWu(09)}, \cite{JungTremayne(11)}, \cite%
{McKenzie(03)}, \cite{ZhuJoe(03)}, \cite{JacobsLewis(78)}, \cite%
{McKenzie(03)} and \cite{Weiss(08)}. Trawl processes also fall within the
wide class of the so-called ambit fields (e.g. \cite%
{BarndorffNielsenSchmiegel(07)} and \cite{BarndorffNielsenBenthVeraart(11)}%
). Recently, \cite{ShephardYang(14)} models high frequency financial data by
using a trawl process to allow for fleeting movements to prices in addition
to an integer-valued L\'{e}vy process proposed by \cite%
{BarndorffNielsenPollardShephard(12)}.

As far as we know, there is no existing literature that directly and
completely addresses likelihood inference for these trawl processes---or
equivalently the prediction based upon it. Even though there are a large
number of papers that focus on likelihood inference for marked point
processes (see \cite{DaleyVere-Jones(08_Ch14)} for a survey), it only
indirectly and partially describes trawl processes in terms of their jumps.
A thorough likelihood inference for trawl processes needs to include the
information in the initial value of the process.

In this Chapter, we provide a thorough analysis of likelihood inference for
integer-valued trawl processes and demonstrate the core
ideas---prediction decomposition, filtering, smoothing and EM algorithm---by
focusing on the so-called exponential trawl. It is not only a simplification
of the modelling framework but also an intellectually interesting special
case of its own, as in this special case the resulting trawl process is a
continuous time hidden Markov process with countable state space. The
theoretical analysis for the filtering and smoothing problems for this type
of process has been discussed in details by \cite{Rudemo(73)} and \cite%
{Rudemo(75)}, using the classical theory of Kolmogorov's forward and
backward differential equations. We particulary emphasize that the resulting
EM algorithm in this special case is exact in the sense that there are no
discretization errors in its computation.

The major goal of this Chapter is to derive filtering and smoothing results
in the framework of trawl processes, so the analysis adopted here can be
easily scaled up to adapt to the discussions of other general trawls or even
the inclusion of a non-stationary component proposed in \cite%
{ShephardYang(14)}. These general discussions will be dealt with elsewhere,
for they require a significantly more sophisticated particle filtering and
smoothing device. We also discuss
non-negative trawl processes, which are particularly easy to work with.

The structure of this Chapter is as follows. In Section \ref{Section: Trawl
Process}, we remind the reader how to construct trawl processes using the
exponential trawl. Section \ref{Section: Fitlering and Smoothing} includes
details of how to carry out filtering and smoothing for these models. In
Section \ref{Section: Likelihood inferences}, we show likelihood inference
for exponential-trawl processes based on these filters and smoothers.
Section \ref{Section: Non-Negative case likelihood} discusses the important
but analytically tractable special case of non-negative trawl processes. We
finally conclude in Section \ref{Section: Conclude}. The Appendix contains
the proofs and derivations of various results given in this Chapter.

\section{Exponential-Trawl Processes\label{Section: Trawl Process}}

In this Section, we build our notation, definitions and key structures for
the exponential-trawl process that will be focused on throughout this
Chapter. We also provide its log-likelihood function based on observed data.

\subsection{Definition}

Our model will be based on a homogeneous L\'{e}vy basis on $\left[ 0,1\right]
\times
\mathbb{R}
\longmapsto
\mathbb{Z}
\backslash \left\{ 0\right\} $, which models the discretely scattered events
of integer size (with direction) $y\in
\mathbb{Z}
\backslash \left\{ 0\right\} $ at each point in time $s\in
\mathbb{R}
$ and height $x\in \left[ 0,1\right] $. It is defined by%
\begin{equation*}
L\left( \mathrm{d}x,\mathrm{d}s\right) \triangleq \int_{-\infty }^{\infty
}yN\left( \mathrm{d}y,\mathrm{d}x,\mathrm{d}s\right) ,\ \ \ \ \left(
x,s\right) \in \left[ 0,1\right] \times
\mathbb{R}
,
\end{equation*}%
where $N$ is a three-dimensional Poisson random measure with intensity
measure%
\begin{equation*}
\mathbb{E}\left( N\left( \mathrm{d}y,\mathrm{d}x,\mathrm{d}s\right) \right)
=\nu \left( \mathrm{d}y\right) \mathrm{d}x\mathrm{d}s.
\end{equation*}%
Here $\mathrm{d}s$ means the arrival times are uniformly scattered (over $%
\mathbb{R}
$), $\mathrm{d}x$ means the random heights are also uniformly scattered
(over $\left[ 0,1\right] $) and $\nu \left( \mathrm{d}y\right) $ is a L\'{e}%
vy measure concentrated on the non-zero integers $%
\mathbb{Z}
\backslash \left\{ 0\right\} $. Without any confusion, we will abuse the
notation $\nu \left( y\right) $ to denote the mass of the L\'{e}vy measure
centered at $y$. Throughout this Chapter, we assume that
\begin{equation*}
\int_{-\infty }^{\infty }\nu \left( \mathrm{d}y\right) =\sum_{y\in
\mathbb{Z}
\backslash \left\{ 0\right\} }\nu \left( y\right) <\infty .
\end{equation*}

Following \cite{BarndorffNielsenLundeShephardVeraart(14)}, we think of
dragging a fixed Borel measurable set $A\subseteq \lbrack 0,1]\times \left(
-\infty ,0\right] $ through time%
\begin{equation*}
A_{t}\triangleq A+(0,t),\ \ \ \ t\geq 0,
\end{equation*}%
so the trawl process is defined by%
\begin{equation*}
Y_{t}\triangleq L(A_{t})=\int_{\left[ 0,1\right] \times
\mathbb{R}
}1_{A}\left( x,s-t\right) L\left( \mathrm{d}x,\mathrm{d}s\right) .
\end{equation*}

Throughout the rest of this Chapter, we will focus on the exponential trawl%
\begin{equation*}
A\triangleq \left\{ \left( x,s\right) :s\leq 0,\ 0\leq x<d\left( s\right)
\triangleq \exp \left( s\phi \right) \right\} ,\ \ \ \ \phi >0,
\end{equation*}%
to simplify our exposition of the key ideas. We will leave results on more
general trawls in another study.

\begin{example}
Suppose that%
\begin{equation*}
\nu (\mathrm{d}y)=\nu ^{+}\delta _{\left\{ 1\right\} }\left( \mathrm{d}%
y\right) +\nu ^{-}\delta _{\left\{ -1\right\} }\left( \mathrm{d}y\right) ,\
\ \ \ \nu ^{+},\nu ^{-}>0,
\end{equation*}%
where $\delta _{\left\{ \pm 1\right\} }\left( \mathrm{d}y\right) $ is the
Dirac point mass measure centered at $\pm 1$. The corresponding $L\left(
\mathrm{d}x,\mathrm{d}s\right) $ is called a Skellam L\'{e}vy basis, while
the special case of $\nu ^{-}=0$ is called Poisson. The upper panel of Fig. %
\ref{Fig.: Trawl process illustration} shows events in $L $ using $\nu
^{+}=\nu ^{-}=10$, taking sizes on $1,-1$ with black and white dots
respectively and with equal probability.
\begin{figure}[t]
\centering\includegraphics[width=11.7cm]{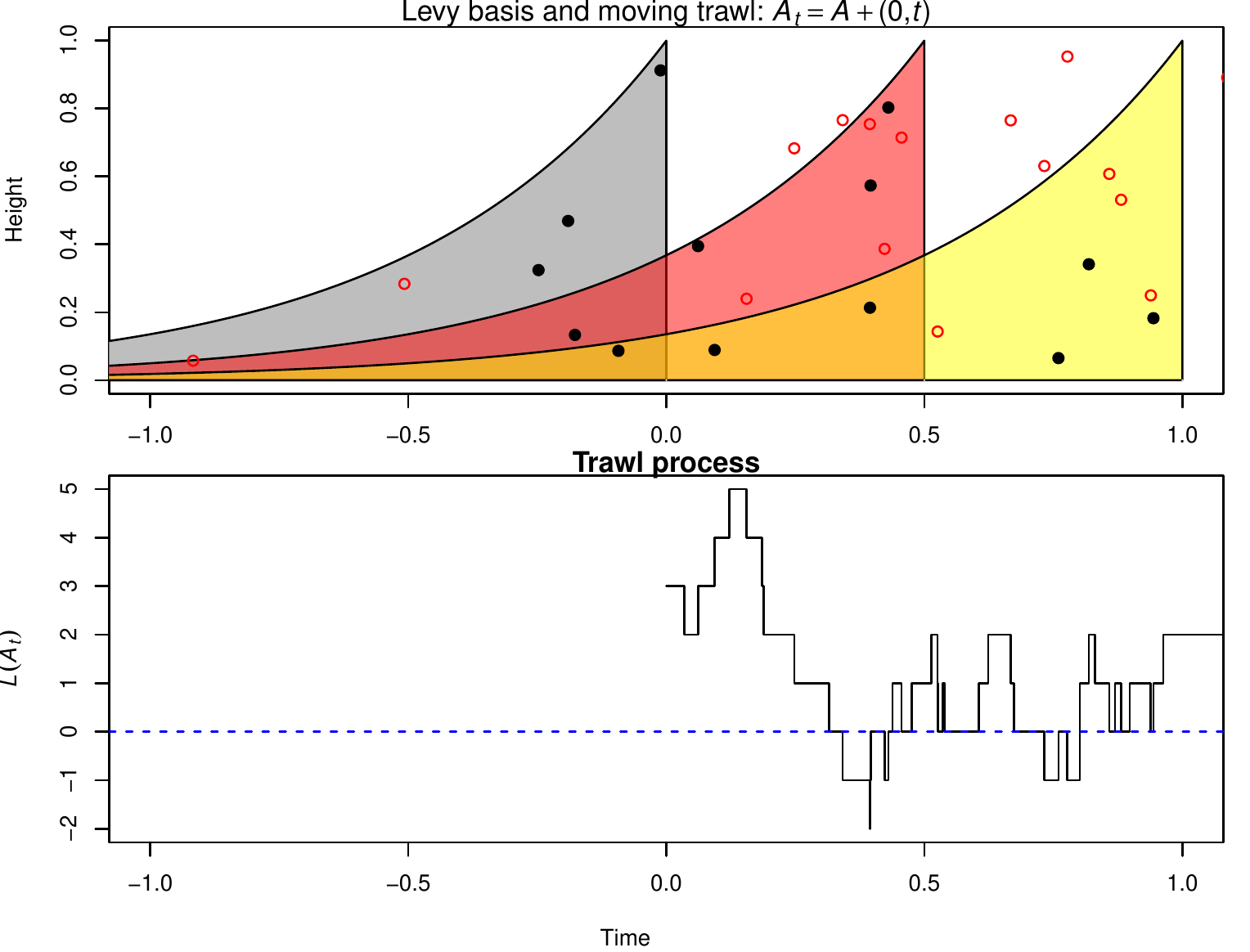}
\caption{A moving trawl $A_{t}$ is joined by the Skellam L\'{e}vy basis $L(%
\mathrm{d}x,\mathrm{d}s)$, where the horizontal axis $s$ is time and the
vertical axis $x$ is height. The shaded area is an example of the
exponential trawl $A$, while we also show the outlines of $A_{t}$ when $%
t=1/2 $ and $t=1$. Also shown below is the implied trawl process $%
Y_{t}=L(A_{t})$. Code: \texttt{EPTprocess\_Illurstration.R}}
\label{Fig.: Trawl process illustration}
\end{figure}
The lower panel of Fig. \ref{Fig.: Trawl process illustration} then
illustrates the resulting Skellam exponential-trawl process $Y_{t}=L\left(
A_{t}\right) $ using $\phi =2$, which sums up all the effects (both positive
and negative) captured by the exponential trawl. Dynamically, $L\left(
A_{t}\right) $ will move up by $1$ if the moving trawl $A_{t}$ either
captures one positive event or releases a negative one; conversely, it will
move down by $1$ if vice versa. Notice that $Y_{0}=L\left( A_{0}\right) $
might not be necessarily zero and the path of $Y$ at negative time is not
observed.
\end{example}

\subsection{Markovian Counting Process\label{Sect: Trawl process
decomposition}}

For $y\in
\mathbb{Z}
\backslash \left\{ 0\right\} $, let $C_{t}^{\left( y\right) }\in \left\{
0,1,2,...\right\} $ be the total counts of surviving events of size $y$ in
the trawl at time $t$, which also includes the event that arrives \emph{%
exactly} at time $t$, so each $C_{t}^{\left( y\right) }$ must be c\`{a}dl\`{a%
}g (right-continuous with left-limits). Then clearly the trawl process can
be represented as%
\begin{equation}
Y_{t}=\sum_{y\in
\mathbb{Z}
\backslash \left\{ 0\right\} }yC_{t}^{(y)},\ \ \ \ t\geq 0.
\label{Trawl process deompose to counting processes}
\end{equation}

Note that each $C_{t}^{\left( y\right) }$ is not only a Poisson
exponential-trawl process with (different) intensity of arrivals $\nu \left(
y\right) $ (and sharing the same trawl) but also a $\mathrm{M}/\mathrm{M}%
/\infty $ queue and hence a continuous time Markov process. Hence, for $%
\left\{ \mathcal{C}_{t}^{\left( y\right) }\right\} _{t\geq 0}$ being the
natural filtration generated by the counting process $C_{t}^{\left( y\right)
}$, i.e., $\mathcal{C}_{t}^{\left( y\right) }\triangleq \sigma \left(
\left\{ C_{s}^{\left( y\right) }\right\} _{0\leq s\leq t}\right) $, it has
(infinitesimal) transition probabilities (or rates or intensities)%
\begin{equation}
\lim\limits_{\mathrm{d}t\rightarrow 0}\dfrac{\mathbb{P}\left( \left.
C_{t}^{(y)}-C_{t-\mathrm{d}t}^{(y)}=j\right\vert \mathcal{C}_{t-\mathrm{d}%
t}^{\left( y\right) }\right) }{\mathrm{d}t}=\left\{
\begin{array}{cc}
\nu \left( y\right) , & \text{if }j=1 \\
\phi C_{t-}^{\left( y\right) }, & \text{if }j=-1 \\
0, & \text{if }j\in
\mathbb{Z}
\backslash \left\{ -1,1\right\}%
\end{array}%
\right. .  \label{Transition intensities of C^(y)}
\end{equation}%
The cases of $j=1$ or $-1$---which correspond to the arrival of a new event
of size $y$ and the departure of an old one---are the only two possible
infinitessimal movements of $C_{t}^{\left( y\right) }$ due to the point
process nature of the L\'{e}vy basis. Note that the arrival
rate and departure rate are controlled by the L\'{e}vy measure $\nu $ and
the trawl parameter $\phi $ respectively. Derivation of (\ref{Transition
intensities of C^(y)}) can be found in many standard references for queue
theory (e.g. \cite{Asmussen(03)}).

\begin{remark}
Let $\Delta X_{t}\triangleq X_{t}-X_{t-}$ denote the \emph{instantaneous}
jump of any process $X$ at time $t$. Then the transition probability (\ref%
{Transition intensities of C^(y)}) can be conveniently written in a
differential form%
\begin{equation*}
\mathbb{P}\left( \left. \Delta C_{t}^{(y)}=j\right\vert \mathcal{C}%
_{t-}^{\left( y\right) }\right) =\left\{
\begin{array}{cc}
\nu \left( y\right) \mathrm{d}t, & \text{if }j=1 \\
\phi C_{t-}^{\left( y\right) }\mathrm{d}t, & \text{if }j=-1 \\
0, & \text{if }j\in
\mathbb{Z}
\backslash \left\{ -1,1\right\}%
\end{array}%
\right. .
\end{equation*}%
Throughout this Chapter, our analysis will be majorly based on this
infinitessimal point of view for the ease of demonstration. All of our
arguments can be rephrased in a mathematically tighter way.
\end{remark}

The independence property of the L\'{e}vy basis implies the independence
between each $C_{t}^{\left( y\right) }$ for $y\in
\mathbb{Z}
\backslash \left\{ 0\right\} $, so the joint count process%
\begin{equation*}
\mathbf{C}_{t}\triangleq \left( ...,C_{t}^{\left( -2\right) },C_{t}^{\left(
-1\right) },C_{t}^{\left( 1\right) },C_{t}^{\left( 2\right) },...\right)
\end{equation*}%
is also Markovian, which serves as the unobserved state process for the
observed \emph{hidden Markov process} $Y_{t}$ and will be the central target
for the filter and smoother we will discuss in a moment. Let $\mathcal{C}%
_{t}\triangleq \sigma \left( \left\{ \mathbf{C}_{s}\right\} _{0\leq s\leq
t}\right) =\dbigvee_{y\in
\mathbb{Z}
\backslash \left\{ 0\right\} }\mathcal{C}_{t}^{\left( y\right) }$ be the
join filtration. Clearly, from (\ref{Transition intensities of C^(y)}), $%
\mathbf{C}_{t}$ has (infinitesimal) transition probabilities%
\begin{equation}
\mathbb{P}\left( \left. \Delta \mathbf{C}_{t}=\mathbf{j}\right\vert \mathcal{%
C}_{t-}\right) =\left\{
\begin{array}{cc}
\nu \left( y\right) \mathrm{d}t, & \text{if }\mathbf{j}=\mathbf{1}^{\left(
y\right) }\text{ for some }y \\
\phi C_{t-}^{\left( y\right) }\mathrm{d}t, & \text{if }\mathbf{j}=-\mathbf{1}%
^{\left( y\right) }\text{ for some }y \\
0, & \text{otherwise}%
\end{array}%
\right. ,  \label{Transition probability of vector C}
\end{equation}%
where $\mathbf{1}^{\left( y\right) }\in
\mathbb{Z}
^{\infty }$ is the vector that takes $1$ at $y$-th component and $0$
otherwise.

The trawl process $Y_{t}$ can be also written as%
\begin{equation*}
Y_{t}=\sum_{y=1}^{\infty }yY_{t}^{(y)},\ \ \ \ Y_{t}^{(y)}\triangleq
C_{t}^{(y)}-C_{t}^{(-y)},
\end{equation*}%
where each $Y_{t}^{(y)}$ is a Skellam exponential-trawl process. Each $%
Y_{t}^{(y)}$ is observed from the path of $Y_{t}$ up to its initial value $%
Y_{0}^{\left( y\right) }$, for we can exactly observe all the jumps of $%
Y_{t} $ and hence allocate them into the appropriate $Y_{t}^{(y)}$. In other
words, we can regard the observed trawl process as (i) a \emph{marked point
process} $\Delta Y_{t}\in
\mathbb{Z}
\backslash \left\{ 0\right\} $, which consists of several independent (given
all the $Y_{0}^{\left( y\right) }$) marked point process $\Delta
Y_{t}^{\left( y\right) }\in \left\{ -1,1\right\} $, plus (ii) the initial
value $Y_{0}$. The missing components $Y_{0}^{\left( y\right) }$'s will have
some mild effects on $\Delta Y_{t}^{\left( y\right) }$. It is this initial
value challenge that differentiates the likelihood analysis of trawl
processes from that of marked point processes.

The special case where $Y_{t}$ is always non-negative has further simpler
structure, as we must have $C_{t}^{\left( -y\right) }=0$ for all $y=1,2,...$
and hence $C_{t}^{\left( y\right) }=Y_{t}^{\left( y\right) }$ is directly
observed up to its initial condition $C_{0}^{\left( y\right) }$, which can
be well-approximated if the observation period $T$ is large enough. We will
go through these details in Section \ref{Section: Non-Negative case
likelihood}.

\subsection{Conditional Intensities and Log-likelihood\label{Section:
Observed data likelihood}}

Let $\left\{ \mathcal{F}_{t}\right\} _{t\geq 0}$ be the natural filtration
generated by the observed trawl process $Y_{t}$, i.e. $\mathcal{F}%
_{t}\triangleq \sigma \left( \left\{ Y_{s}\right\} _{0\leq s\leq t}\right) $%
. Define the c\`{a}dl\`{a}g conditional intensity process of the trawl
process $Y$ as%
\begin{equation}
\lambda _{t-}^{\left( y\right) }\triangleq \lim\limits_{\mathrm{d}%
t\rightarrow 0}\dfrac{\mathbb{P}\left( Y_{t}-Y_{t-\mathrm{d}t}=y|\mathcal{F}%
_{t-\mathrm{d}t}\right) }{\mathrm{d}t},\ \ \ \ y\in
\mathbb{Z}
\backslash \left\{ 0\right\} ,\ t>0
\label{Definition of conditional intensity}
\end{equation}%
or conveniently in a differential form%
\begin{equation}
\lambda _{t-}^{\left( y\right) }\mathrm{d}t\triangleq \mathbb{P}\left(
\Delta Y_{t}=y|\mathcal{F}_{t-}\right) .
\label{Diff. Def. of conditional intensity}
\end{equation}%
It means the (time-varying) predictive intensity of a size $y$ move at time $%
t$ of the trawl process, conditional on information instantaneously before
time $t$.

\begin{remark}
To emphasize the $\mathcal{F}_{t}$-predictability of $\lambda ^{\left(
y\right) }$, i.e., being adapted to the left natural filtration $\mathcal{F}%
_{t-}$, we will keep the subscript $t-$ throughout this Chapter. This is
particularly informative in the implementation of likelihood
calculations, reminding us to take the \emph{left-limit} of the
intensity process whenever there is a jump.
\end{remark}

For any two $\sigma $-fields $\mathcal{F}$ and $\mathcal{G}$, let the
Radon-Nikodym derivative over $\mathcal{F}|\mathcal{G}$ between two
probability measures $\mathbb{P}$ and $%
\mathbb{Q}
$ be%
\begin{equation*}
\left( \dfrac{\mathrm{d}\mathbb{\mathbb{P}}}{\mathrm{d}%
\mathbb{Q}
}\right) _{\mathcal{F}|\mathcal{G}}\triangleq \dfrac{\left( \dfrac{\mathrm{d}%
\mathbb{\mathbb{P}}}{\mathrm{d}%
\mathbb{Q}
}\right) _{\mathcal{F}\vee \mathcal{G}}}{\left( \dfrac{\mathrm{d}\mathbb{%
\mathbb{P}}}{\mathrm{d}%
\mathbb{Q}
}\right) _{\mathcal{G}}}.
\end{equation*}%
In particular, when $\mathcal{G}=\sigma \left( X\right) $ for any random
variable $X$, we will simply write the subscript as $\mathcal{F}|X$. The
following classical result serves as the foundation for all likelihood
inference for jump processes.

\begin{theorem}
\label{Thm.: Point process Radon-Nykodym Derivative}Let $X_{t}$ be any
integer-valued stochastic process and $\left\{ \mathcal{F}_{t}^{X} \right\}
_{t \geq 0}$ be its associated natural filtration. Assume that, under both $%
\mathbb{P}$ and $%
\mathbb{Q}
$, (i) it has finite expected number of jumps during $(0,T]$, and (ii) the
conditional intensities $\lambda _{t-}^{\left( y\right) ,\mathbb{P}}$ and $%
\lambda _{t-}^{\left( y\right) ,%
\mathbb{Q}
}$ are well-defined using (\ref{Definition of conditional intensity}) and $%
\mathcal{F}_{t-}^{X}$. Then $\mathbb{P\ll
\mathbb{Q}
}$ over $\mathcal{F}_{T}^{X}|X_{0}$ if and only if $\lambda _{t-}^{\left(
y\right) ,%
\mathbb{Q}
}$ is strictly positive. In this case, the logarithmic Radon-Nikodym
derivative over $\mathcal{F}_{T}^{X}|X_{0}$ is%
\begin{eqnarray*}
\log \left( \dfrac{\mathrm{d}\mathbb{P}}{\mathrm{d}\mathbb{%
\mathbb{Q}
}}\right) _{\mathcal{F}_{T}^{X}|X_{0}} &=&\sum_{0<t\leq T}\sum_{y\in
\mathbb{Z}
\backslash \left\{ 0\right\} }\log \left( \dfrac{\lambda _{t-}^{\left(
y\right) ,\mathbb{P}}}{\lambda _{t-}^{\left( y\right) ,%
\mathbb{Q}
}}\right) 1_{\left\{ \Delta X_{t}=y\right\} } \\
&&-\int_{t\in (0,T]}\sum_{y\in
\mathbb{Z}
\backslash \left\{ 0\right\} }\left( \lambda _{t-}^{\left( y\right) ,\mathbb{%
P}}-\lambda _{t-}^{\left( y\right) ,\mathbb{%
\mathbb{Q}
}}\right) \mathrm{d}t.
\end{eqnarray*}
\end{theorem}

Proposition 14.4.I of \cite{DaleyVere-Jones(08_Ch14)} provides a complete
and mathematically rigorous treatment for this Theorem. For completeness, we
also provide an intuitive and heuristic derivation in the Appendix. A direct
application of Theorem \ref{Thm.: Point process Radon-Nykodym Derivative}
gives the following Corollary.

\begin{corollary}
\label{Cor.: log-likelihood for general trawl processes}The log-likelihood
function of the (general) trawl process is (ignoring the constant)%
\begin{equation}
l_{\mathcal{F}_{T}}\left( \mathbf{\theta }\right) =\sum_{0<t\leq
T}\sum_{y\in
\mathbb{Z}
\backslash \left\{ 0\right\} }\log \lambda _{t-}^{\left( y\right)
}1_{\left\{ \Delta Y_{t}=y\right\} }-\int_{t\in (0,T]}\sum_{y\in
\mathbb{Z}
\backslash \left\{ 0\right\} }\lambda _{t-}^{\left( y\right) }\mathrm{d}%
t+l_{Y_{0}}\left( \mathbf{\theta }\right) ,  \label{Log-Likelihood}
\end{equation}%
where the parameters of interest $\mathbf{\theta }$ include the L\'{e}vy
measure $\nu \left( \mathrm{d}y\right) $ (i.e. $\nu \left( y\right) $'s) and
the trawl parameter $\phi $.
\end{corollary}

The study of likelihood inference for trawl processes then reduces to the
calculations of conditional intensities $\lambda _{t-}^{\left( y\right) }$
for $y\in
\mathbb{Z}
\backslash \left\{ 0\right\} $. Now, by law of iterated expectations and the
fact that $\mathcal{C}_{t}\supseteq \mathcal{F}_{t}$ for all $t$ (because of
(\ref{Trawl process deompose to counting processes})), we have%
\begin{eqnarray*}
\lambda _{t-}^{\left( y\right) }\mathrm{d}t &=&\mathbb{E}\left( \mathbb{P}%
\left( \Delta Y_{t}=y|\mathcal{C}_{t-}\right) |\mathcal{F}_{t-}\right) \\
&=&\mathbb{E}\left( \left. \mathbb{P}\left( \left. \Delta \mathbf{C}_{t}=%
\mathbf{1}^{\left( y\right) }\right\vert \mathcal{C}_{t-}\right) \right\vert
\mathcal{F}_{t-}\right) +\mathbb{E}\left( \left. \mathbb{P}\left( \left.
\Delta \mathbf{C}_{t}=-\mathbf{1}^{\left( -y\right) }\right\vert \mathcal{C}%
_{t-}\right) \right\vert \mathcal{F}_{t-}\right) \\
&=&\nu \left( y\right) \mathrm{d}t+\phi \mathbb{E}\left( \left.
C_{t-}^{\left( -y\right) }\right\vert \mathcal{F}_{t-}\right) \mathrm{d}t,
\end{eqnarray*}%
where the second line follows because the event $\Delta Y_{t}=y$ must come
from either an arrival of a new size $y$ event or a departure of an old size
$-y$ event; the third line follows from (\ref{Transition probability of
vector C}). Thus,%
\begin{equation}
\lambda _{t-}^{\left( y\right) }=\nu \left( y\right) +\phi \mathbb{E}\left(
\left. C_{t-}^{\left( -y\right) }\right\vert \mathcal{F}_{t-}\right) ,\ \ \
\ y\in
\mathbb{Z}
\backslash \left\{ 0\right\} .  \label{conditional intensity}
\end{equation}%
In next Section, we will study an exact filtering scheme to numerically
calculate $\mathbb{E}\left( \left. C_{t-}^{\left( -y\right) }\right\vert
\mathcal{F}_{t-}\right) $.

The non-negative exponential-trawl process, where we always have positive
events, admits a further simplification%
\begin{equation}
\lambda _{t-}^{\left( y\right) }=\nu \left( y\right) ,\ \ \ \ \lambda
_{t-}^{\left( -y\right) }=\phi \mathbb{E}\left( \left.
C_{t-}^{(y)}\right\vert \mathcal{F}_{t-}\right) ,\ \ \ \ y=1,2,...,
\label{exponential trawl non-negative}
\end{equation}%
so likelihood inference for such a case is easier. In the Poisson case, all
the impacts are of size one, so in particular $C_{0}^{(1)}=Y_{0}$ is also
observed (as $C_{0}^{\left( y\right) }=0$ for all $y\neq 1$), which allows
us to bypass the conditional expectation in (\ref{exponential trawl
non-negative}) for $y=1$.

\section{Exact Filter and Smoother for Exponential-Trawl Processes\label%
{Section: Fitlering and Smoothing}}

\subsection{Filtering\label{Section: Filtering for exponential trawl}}

In general we need to solve the filtering problems for $\mathbf{C}_{t}$ to
implement (\ref{Log-Likelihood}) and (\ref{conditional intensity}). Denote
the filtering probability mass function as%
\begin{equation*}
p_{t,s}\left( \mathbf{j}\right) \triangleq \mathbb{P}\left( \left. \mathbf{C}%
_{t}=\mathbf{j}\right\vert \mathcal{F}_{s}\right) ,\ \ \ \ \mathbf{j}=\left(
...,j_{-2},j_{-1},j_{1},j_{2},...\right) ,\ j_{y}=0,1,2,...,\ t,s\geq 0.
\end{equation*}%
Also, let $\left\Vert \mathbf{j}\right\Vert _{1}\triangleq \sum_{y\in
\mathbb{Z}
\backslash \left\{ 0\right\} }j_{y}$ and $D_{t}\triangleq \left\Vert \mathbf{%
C}_{t}\right\Vert _{1}=\sum_{y\in
\mathbb{Z}
\backslash \left\{ 0\right\} }C_{t}^{\left( y\right) }$.

Our goal here is to sequentially update $p_{t-,t-}\left( \mathbf{j}\right) $%
, where the initial distribution is derived from%
\begin{equation*}
C_{0}^{\left( y\right) }\overset{\text{indep.}}{\backsim }\mathrm{Poisson}%
\left( \dfrac{\nu \left( y\right) }{\phi }\right) \text{ subject to }%
\sum_{y\in
\mathbb{Z}
\backslash \left\{ 0\right\} }yC_{0}^{\left( y\right) }=Y_{0},
\end{equation*}%
so, by letting $\mathrm{Poisson}\left( x|\lambda \right) \triangleq \lambda
^{x}e^{-\lambda }/x!$, we have%
\begin{equation*}
p_{0,0}\left( \mathbf{j}\right) =\dfrac{\prod_{y\in
\mathbb{Z}
\backslash \left\{ 0\right\} }\mathrm{Poisson}\left( j_{y}|\nu \left(
y\right) /\phi \right) }{\mathbb{P}\left( \sum_{y\in
\mathbb{Z}
\backslash \left\{ 0\right\} }yC_{0}^{\left( y\right) }=Y_{0}\right) },
\end{equation*}%
where the denominator can be numerically calculated using the inverse fast
Fourier transform \cite{ShephardYang(14)}.

Notice that the filtering distribution not only updates at the times when
the process jumps but also at those inactivity periods. We discuss these two
cases separately.

\begin{theorem}[Forward Filtering]
\label{Thm.: Filtering}

\begin{enumerate}
\item \lbrack \textbf{Update by inactivity}] Assume that the last jump time
is $\tau $ (or $\tau =0$) and the current time is $t-$, where $\Delta
Y_{s}=0 $ for $\tau <s<t$ (and $\Delta Y_{\tau }\neq 0$ if $\tau >0$). Then%
\begin{equation}
p_{t-,t-}\left( \mathbf{j}\right) =\dfrac{e^{-\phi \left\Vert \mathbf{j}%
\right\Vert _{1}\left( t-\tau \right) }p_{\tau ,\tau }\left( \mathbf{j}%
\right) }{\sum_{\mathbf{k}}e^{-\phi \left\Vert \mathbf{k}\right\Vert
_{1}\left( t-\tau \right) }p_{\tau ,\tau }\left( \mathbf{k}\right) },
\label{Filtering update by inactivity}
\end{equation}%
where $p_{\tau ,\tau }$ is the filtering distribution we have already known
at time $\tau $.

\item \lbrack \textbf{Update by jump}] Assume that the current time is $\tau
-$ and $\Delta Y_{\tau }=y$ for some $y\in
\mathbb{Z}
\backslash \left\{ 0\right\} $. Then%
\begin{equation}
p_{\tau ,\tau }\left( \mathbf{j}\right) =\dfrac{1}{\lambda _{\tau -}^{\left(
y\right) }}\left( \nu \left( y\right) p_{\tau -,\tau -}\left( \mathbf{j}-%
\mathbf{1}^{\left( y\right) }\right) +\phi \left( j_{-y}+1\right) p_{\tau
-,\tau -}\left( \mathbf{j}+\mathbf{1}^{\left( -y\right) }\right) \right) ,
\label{Filtering update by jumps}
\end{equation}%
where $p_{\tau -,\tau -}$ is the filtering distribution we have already
known at time $\tau -$.
\end{enumerate}
\end{theorem}

Overall, the filtering procedures (\ref{Filtering update by inactivity}) and
(\ref{Filtering update by jumps}) imply that $p_{t-,t-}\left( \mathbf{j}%
\right) $ can be updated in continuous time without discretization errors at
any set of finite discrete time points, so we call it an exact filter.

\begin{example}
\label{Ex.: Skellam Filtering}For Skellam exponential-trawl process with L%
\'{e}vy intensities $\nu ^{+}$ and $\nu ^{-}$, we always have%
\begin{equation*}
Y_{t-}=C_{t-}^{\left( +\right) }-C_{t-}^{\left( -\right) },\ \ \ \ t>0,
\end{equation*}%
so knowing $p_{t-,t-}\left( j\right) \triangleq \mathbb{P}\left( \left.
C_{t-}^{\left( -\right) }=j\right\vert \mathcal{F}_{t-}\right) $ immediately
gives us $p_{t-,t-}\left( j,k\right) $. Hence, the filtering updating scheme
reduces to the following: starting from $\tau =0$,%
\begin{eqnarray*}
p_{t-,t-}\left( j\right) &\propto &e^{-\phi \left( 2j+Y_{\tau }\right)
\left( t-\tau \right) }p_{\tau ,\tau }\left( j\right) \ \ \ \ \text{if\ }%
\Delta Y_{s}=0\text{ for }\tau <s<t, \\
p_{\tau ,\tau }\left( j\right) &\propto &\nu ^{+}p_{\tau -,\tau -}\left(
j\right) +\phi \left( j+1\right) p_{\tau -,\tau -}\left( j+1\right) \text{\
\ \ \ if }\Delta Y_{\tau }=1, \\
p_{\tau ,\tau }\left( j\right) &\propto &\nu ^{-}p_{\tau -,\tau -}\left(
j-1\right) +\phi \left( j+Y_{\tau -}\right) p_{\tau -,\tau -}\left( j\right)
\text{\ \ \ \ if }\Delta Y_{\tau }=-1.
\end{eqnarray*}%
We then renormalize $p_{t-,t-}\left( j\right) $ such that $%
\sum_{j=0}^{\infty }p_{t-,t-}\left( j\right) =1$ in each step of the
updates. Knowing the filtering distributions $p_{t-,t-}\left( j\right) $
allows us to calculate%
\begin{equation*}
\mathbb{E}\left( \left. C_{t-}^{\left( -\right) }\right\vert \mathcal{F}%
_{t-}\right) =\sum_{j=0}^{\infty }jp_{t-,t-}\left( j\right) ,\ \mathbb{E}%
\left( \left. C_{t-}^{\left( +\right) }\right\vert \mathcal{F}_{t-}\right)
=\sum_{j=0}^{\infty }jp_{t-,t-}\left( j\right) +Y_{t-}.
\end{equation*}%
Using the following settings, with time unit being second,%
\begin{equation}
\nu ^{+}=0.013,\ \nu ^{-}=0.011,\ \phi =0.034,\ T=21\times 60^{2}=75,600%
\text{\ (sec.)},  \label{Simulated Skellam OU True Setting}
\end{equation}%
Fig. \ref{Fig.: Skellam Filtering} shows a simulated path of the trawl
process $Y_{t}$ together with the filtering expectations of $C_{t}^{\left(
+\right) }$, $C_{t}^{\left( -\right) }$ and $D_{t}=C_{t}^{\left( +\right)
}+C_{t}^{\left( -\right) }$, the total number of surviving (both positive
and negative) events in the trawl at time $t$.
\begin{figure}[t]
\centering%
\includegraphics[width=11.7cm]{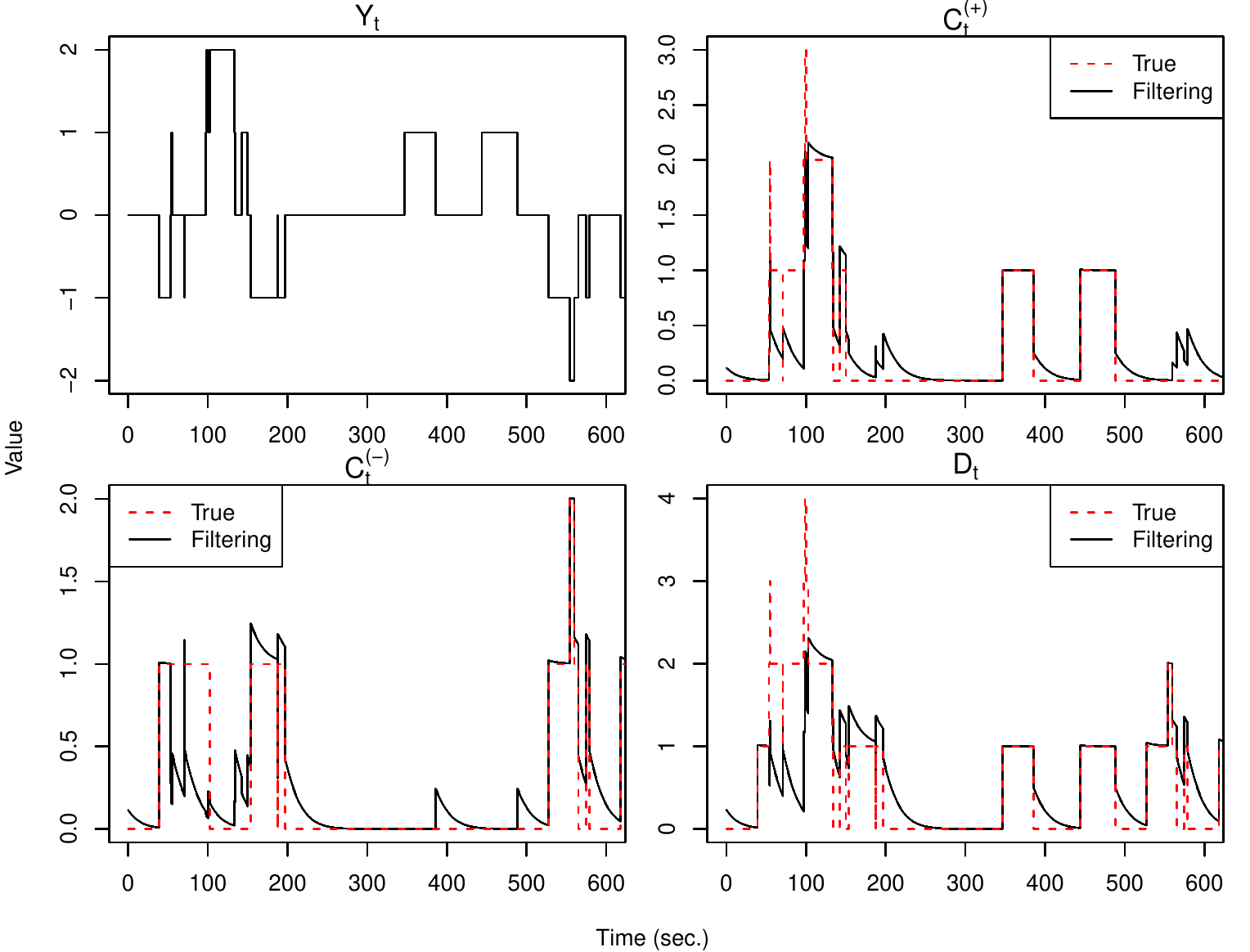}
\caption{\emph{Top left}: A simulated path for the Skellam exponential-trawl
process $Y_{t}$. \emph{Top right}, \emph{Bottom left}, \emph{Bottom right}:
Paths of the true hidden counting processes $C_{t}^{\left( +\right) }$, $%
C_{t}^{\left( -\right) }$ and $D_{t}=C_{t}^{\left( +\right) }+C_{t}^{\left(
-\right) }$ of surviving events in the trawl along with their filtering
estimations. Code: \texttt{EPTprocess\_FilteringSmoothing\_Illustration.R}}
\label{Fig.: Skellam Filtering}
\end{figure}
\end{example}

\subsection{Smoothing}

We now consider the smoothing procedure for the exponential-trawl process $%
Y_{t}$, which is necessary for the likelihood inference based on the EM
algorithm we will see in a moment.

Running the filtering procedure up to time $T$, we then start from $p_{T,T}$
to conduct the smoothing procedure.

\begin{theorem}[Backward Smoothing]
\label{Thm.: Smoothing}

\begin{enumerate}
\item \lbrack \textbf{Update by inactivity}] Assume that the (backward) last
jump time is $\tau $ (or $\tau =T$) and the current time is $t$, where $%
\Delta Y_{s}=0$ for $t\leq s<\tau $ (and $\Delta Y_{\tau }\neq 0$ if $\tau
<T $). Then%
\begin{equation*}
p_{t,T}\left( \mathbf{j}\right) =p_{\tau -,T}\left( \mathbf{j}\right) ,
\end{equation*}%
where $p_{\tau -,T}$ is the smoothing distribution we have already known at
time $\tau -$.

\item \lbrack \textbf{Update by jump}] Assume that the current time is $\tau
$ and $\Delta Y_{\tau }=y$ for some $y\in
\mathbb{Z}
\backslash \left\{ 0\right\} $. Then%
\begin{equation}
p_{\tau -,T}\left( \mathbf{j}\right) =\dfrac{p_{\tau -,\tau -}\left( \mathbf{%
j}\right) }{\lambda _{\tau -}^{\left( y\right) }}\left( \nu \left( y\right)
\dfrac{p_{\tau ,T}\left( \mathbf{j}+\mathbf{1}^{\left( y\right) }\right) }{%
p_{\tau ,\tau }\left( \mathbf{j}+\mathbf{1}^{\left( y\right) }\right) }+\phi
j_{-y}\dfrac{p_{\tau ,T}\left( \mathbf{j}-\mathbf{1}^{\left( -y\right)
}\right) }{p_{\tau ,\tau }\left( \mathbf{j}-\mathbf{1}^{\left( -y\right)
}\right) }\right) ,  \label{Smoothing updating over jump}
\end{equation}%
where $p_{\tau -,\tau -}$ and $p_{\tau ,\tau }$ are from the forward
filtering procedure and $p_{\tau ,T}$ is the smoothing distribution we have
already known at time $\tau $.
\end{enumerate}
\end{theorem}

The two terms in (\ref{Smoothing updating over jump}) are
\begin{equation*}
\mathbb{P}\left( \left. \mathbf{C}_{\tau -}=\mathbf{j},\mathbf{C}_{\tau }=%
\mathbf{j}+\mathbf{1}^{\left( y\right) }\right\vert \mathcal{F}_{T}\right)
\text{ and }\mathbb{P}\left( \left. \mathbf{C}_{\tau -}=\mathbf{j},\mathbf{C}%
_{\tau }=\mathbf{j}-\mathbf{1}^{\left( y\right) }\right\vert \mathcal{F}%
_{T}\right)
\end{equation*}%
respectively, so, in particular,%
\begin{eqnarray}
\mathbb{P}\left( \left. \Delta C_{\tau }^{\left( y\right) }=1\right\vert
\mathcal{F}_{T}\right) &=&\sum_{\mathbf{j}}\dfrac{p_{\tau -,\tau -}\left(
\mathbf{j}\right) }{\lambda _{\tau -}^{\left( y\right) }}\left( \nu \left(
y\right) \dfrac{p_{\tau ,T}\left( \mathbf{j}+\mathbf{1}^{\left( y\right)
}\right) }{p_{\tau ,\tau }\left( \mathbf{j}+\mathbf{1}^{\left( y\right)
}\right) }\right) ,  \label{Smoothing probability of arrival} \\
\mathbb{P}\left( \left. \Delta C_{\tau }^{\left( y\right) }=-1\right\vert
\mathcal{F}_{T}\right) &=&\sum_{\mathbf{j}}\dfrac{p_{\tau -,\tau -}\left(
\mathbf{j}\right) }{\lambda _{\tau -}^{\left( y\right) }}\left( \phi j_{-y}%
\dfrac{p_{\tau ,T}\left( \mathbf{j}-\mathbf{1}^{\left( -y\right) }\right) }{%
p_{\tau ,\tau }\left( \mathbf{j}-\mathbf{1}^{\left( -y\right) }\right) }%
\right) .  \label{Smoothing probability of departure}
\end{eqnarray}%
These (total) weights in (\ref{Smoothing updating over jump}) will be
recorded for every jump time $\tau $ as by-products of the smoothing
procedure, for later they will play important roles in the EM algorithm
introduced in Subsection \ref{Section: EM Algorithm}.

\begin{example}[Continued from Example \protect\ref{Ex.: Skellam Filtering}]

For Skellam exponential-trawl process, the smoothing updating scheme reduces
to the following: starting from $\tau =T$,%
\begin{eqnarray*}
p_{t,T}\left( j\right) &=&p_{\tau -,T}\left( j\right) \ \ \ \ \text{if\ }%
\Delta Y_{s}=0\text{ for }t\leq s<\tau , \\
p_{\tau -,T}\left( j\right) &\propto &p_{\tau -,\tau -}\left( j\right)
\left( \nu ^{+}\dfrac{p_{\tau ,T}\left( j\right) }{p_{\tau ,\tau }\left(
j\right) }+\phi j\dfrac{p_{\tau ,T}\left( j-1\right) }{p_{\tau ,\tau }\left(
j-1\right) }\right) \text{\ \ \ \ if }\Delta Y_{\tau }=1, \\
p_{\tau -,T}\left( j\right) &\propto &p_{\tau -,\tau -}\left( j\right)
\left( \nu ^{-}\dfrac{p_{\tau ,T}\left( j+1\right) }{p_{\tau ,\tau }\left(
j+1\right) }+\phi \left( Y_{\tau -}+j\right) \dfrac{p_{\tau ,T}\left(
j\right) }{p_{\tau ,\tau }\left( j\right) }\right) \text{\ \ \ \ if }\Delta
Y_{\tau }=-1.
\end{eqnarray*}%
We also renormalize $p_{t,T}\left( j\right) $ in each step of the updates.

Using the same simulated path and the same setting (\ref{Simulated Skellam
OU True Setting}) as in Example \ref{Ex.: Skellam Filtering}, we show the
smoothing expectations of $C_{t}^{\left( +\right) }$, $C_{t}^{\left(
-\right) }$ and $D_{t}$ in Fig. \ref{Fig.: Skellam Smoothing}. For most of
the time, the smoothing expectations can match the truth quite well and will
remove those peaks of filtering expectations resulted from departures (such
as the one close to $t=400$ in the plot for $C_{t}^{\left( -\right) }$).
\begin{figure}[th]
\centering%
\includegraphics[width=11.7cm]{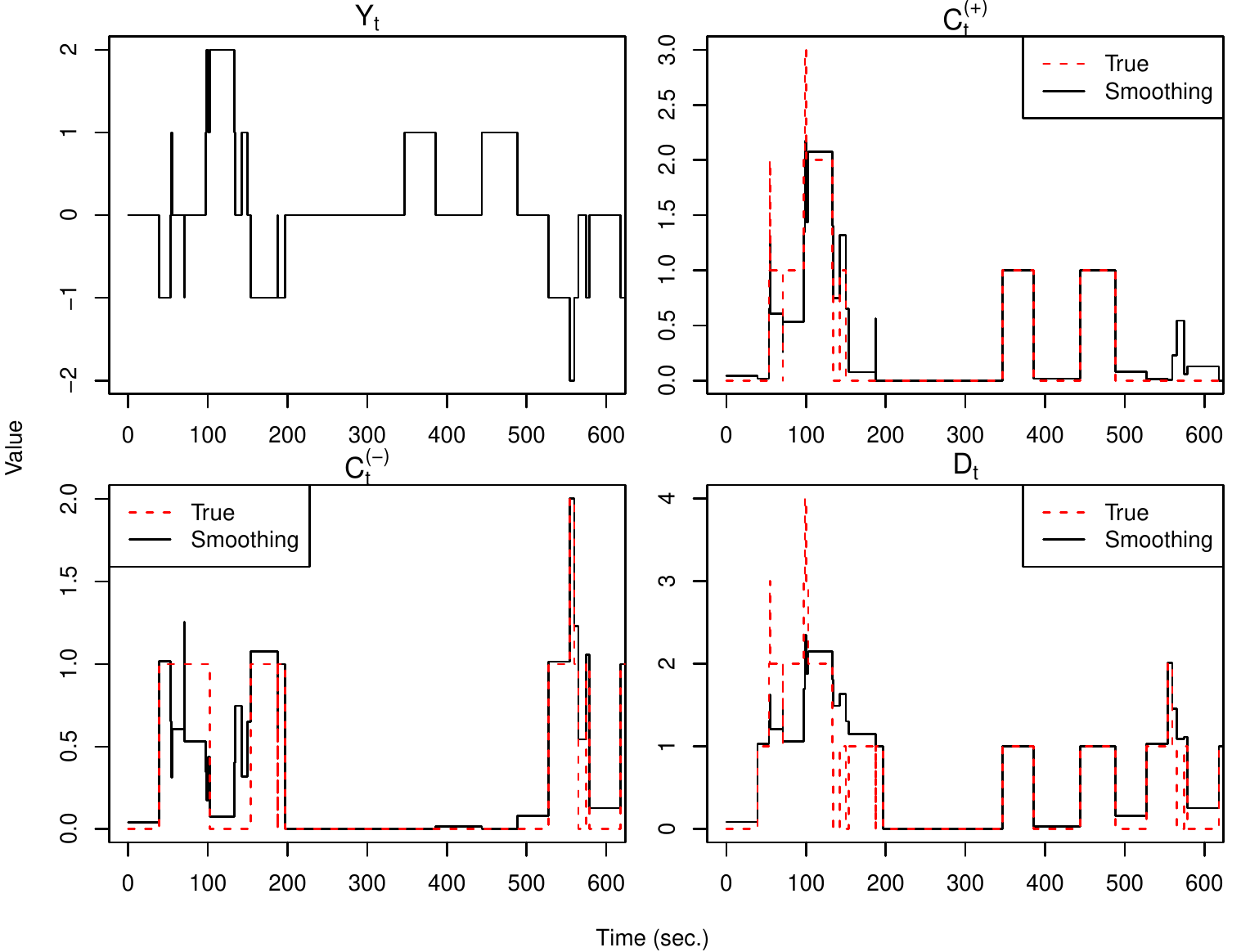}
\caption{\emph{Top left}: A simulated path for the Skellam exponential-trawl
process $Y_{t}$. \emph{Top right}, \emph{Bottom left}, \emph{Bottom right}:
Paths of the true hidden counting processes $C_{t}^{\left( +\right) }$, $%
C_{t}^{\left( -\right) }$ and $D_{t}=C_{t}^{\left( +\right) }+C_{t}^{\left(
-\right) }$ of surviving events in the trawl along with their smoothing
estimations. Code: \texttt{EPTprocess\_FilteringSmoothing\_Illustration.R}}
\label{Fig.: Skellam Smoothing}
\end{figure}
\end{example}

Now we are capable of conducting likelihood inference for exponential-trawl
processes as one of the most important applications of the filtering and
smoothing procedures we have already built here.

\section{Likelihood Inference for General Exponential-Trawl Processes\label%
{Section: Likelihood inferences}}

It has been reported by \cite{BarndorffNielsenLundeShephardVeraart(14)} and
\cite{ShephardYang(14)} that the moment-based inference for the family of
trawl processes could be easily performed, but such inference is arbitrarily
dependent on its procedure design. In this Section, we focus on the maximum likelihood
estimate (MLE) calculation for exponential-trawl processes with general L%
\'{e}vy basis and demonstrate its correctness using several examples.

\subsection{MLE Calculation based on Filtering}

Recall that the evaluation of the log-likelihood (\ref{Log-Likelihood})
requires the calculations of the conditional intensities $\lambda
_{t-}^{\left( y\right) }$ and their integrals%
\begin{equation}
\int_{t\in (0,T]}\lambda _{t-}^{\left( y\right) }\mathrm{d}t=\nu \left(
y\right) T+\phi \sum_{\mathbf{j}}j_{-y}\int_{t\in (0,T]}p_{t-,t-}\left(
\mathbf{j}\right) \mathrm{d}t,  \label{Intensity integral}
\end{equation}%
which follows from (\ref{conditional intensity}) and $\mathbb{E}\left(
\left. C_{t-}^{\left( -y\right) }\right\vert \mathcal{F}_{t-}\right) =\sum_{%
\mathbf{j}}j_{-y}p_{t-,t-}\left( \mathbf{j}\right) $.

However, we do not know the integral $\int_{t\in (0,T]}p_{t-,t-}\left(
\mathbf{j}\right) \mathrm{d}t$ analytically, as the denominator in (\ref%
{Filtering update by inactivity}) also depends on $t$. Hence, we have to
calculate (\ref{Filtering update by inactivity}) in a dense grid of time
points---separated by a time gap $\delta _{\mathrm{inactivity}}$ during
those inactivity periods---and approximate (\ref{Intensity integral}) by
linear interpolation. Clearly, the smaller the time gap $\delta _{\mathrm{%
inactivity}}$, the smaller the numerical error in (\ref{Intensity integral})
but the larger the computational burden.

\begin{example}
\label{Ex.: MLE FIltering large vs small delta}Using the true parameters in (%
\ref{Simulated Skellam OU True Setting}) and simulating a $10$-day-long data
with $T=756,000$ (sec.), Fig. \ref{Fig.: Comparison btw. different delta}
shows how an inappropriate choice of $\delta _{\mathrm{inactivity}}$ will
depict a wrong log-likelihood surface no matter how long the correct
simulated data we supply, where the comparison is made with respect to the
first day portion ($75,600$ (sec.)) of the $10$-day-long simulated data.
\begin{figure}[th]
\centering%
\includegraphics[width=11.7cm]{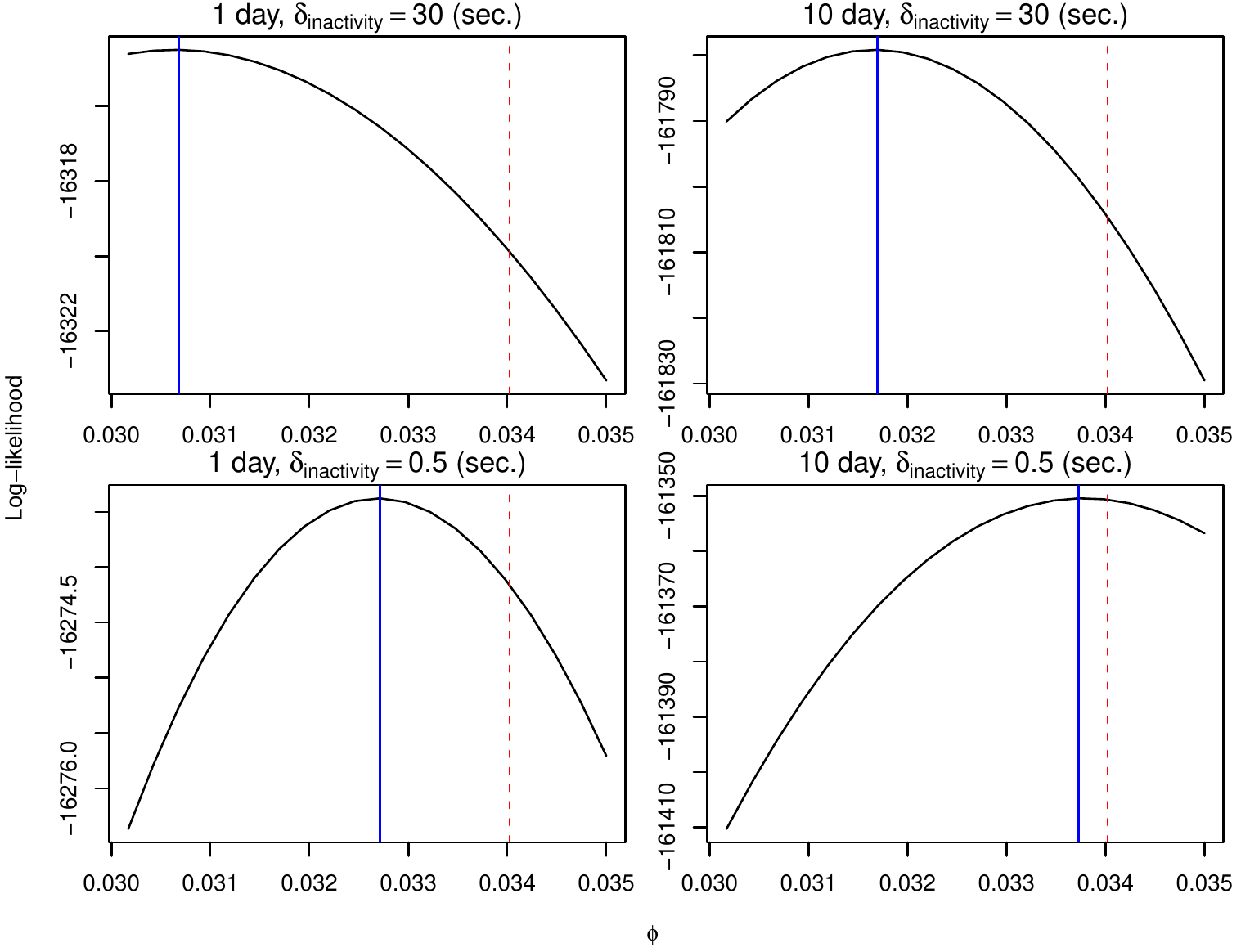}
\caption{Log-likelihood plots over $\protect\phi $ (with $\protect\nu ^{+}$
and $\protect\nu ^{-}$ fixed at the truth) using different $\protect\delta _{%
\mathrm{inactivity}}$ and a simulated $10$-day-long ($T=756,000$ (sec.))
Skellam exponential-trawl process. The one-day-long data is the first tenth
of the simulated data. The dashed lines indicate the true value of $\protect%
\phi $, while the solid lines indicate the optimal value of $\protect\phi $
in each plot. The $p$-values using the likelihood ratio test are $0.104\%$ (%
\emph{Top left}), $21.0\%$ (\emph{Bottom left}), $8.82\times 10^{-13}$ (%
\emph{Top right}) and $46.1\%$ (\emph{Bottom right}). Code: \texttt{%
EPTprocess\_MLE\_Inference\_Simulation\_Small\_vs\_Large.R}}
\label{Fig.: Comparison btw. different delta}
\end{figure}
Using the same one-day-long data, Fig. \ref{Fig.: Log-likelihood plots over
intensities} also shows the corresponding log-likelihood function over $\nu
^{+}$ or $\nu ^{-}$ with other parameters fixed at the truth. Including the
bottom left panel of Fig. \ref{Fig.: Comparison btw. different delta}, all
of the MLE's (solid lines) are reasonably close to the true values (dashed
lines) and the likelihood ratio tests suggest that $p$-values are all
greater than $20\%$.
\begin{figure}[th]
\centering%
\includegraphics[width=11.7cm]{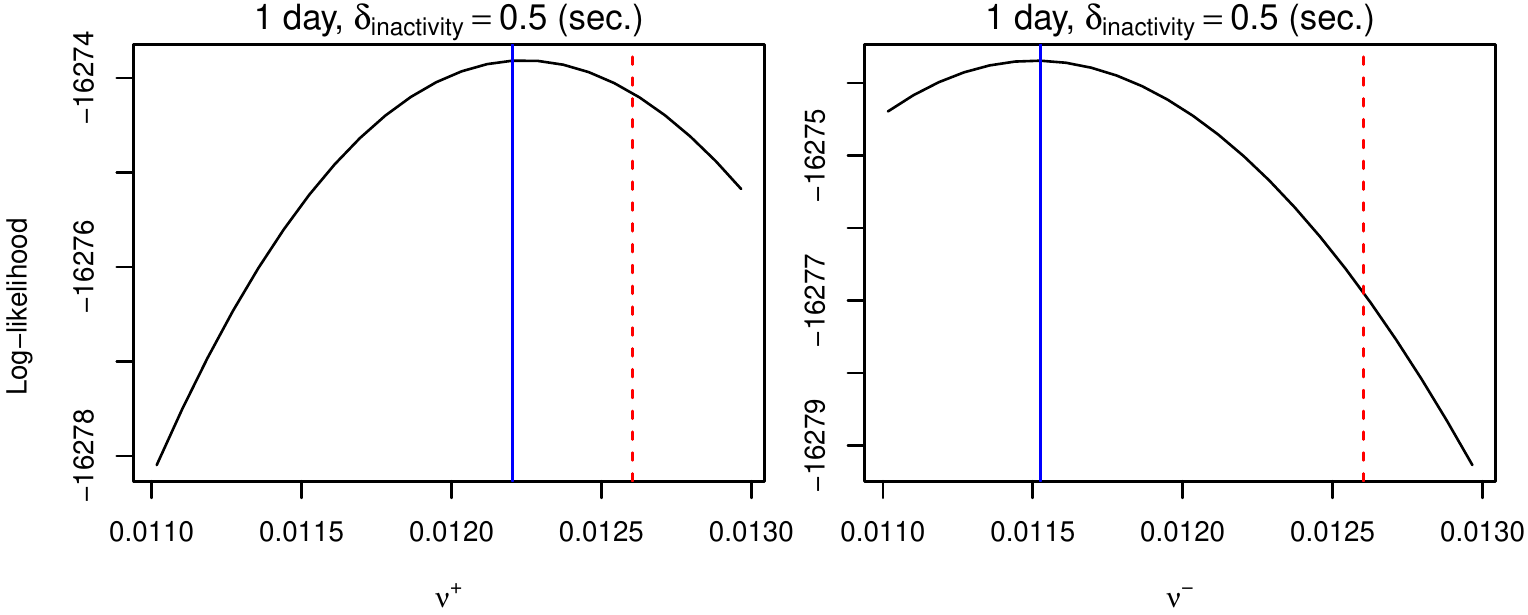}
\caption{Log-likelihood plots over either $\protect\nu ^{+}$ or $\protect\nu %
^{-}$ for one simulated Skellam exponential-trawl process. The dashed lines
indicate the true value, while the solid lines indicate the optimal value of
$\protect\nu ^{+}$ or $\protect\nu ^{-}$ in the individual plot. The $p$%
-values using the likelihood ratio test are $40.5\%$ (\emph{Left}) and $%
33.4\%$ (\emph{Right}). Code: \texttt{EPTprocess\_MLE\_Inference\_Simulation%
\_Small\_vs\_Large.R}}
\label{Fig.: Log-likelihood plots over intensities}
\end{figure}
\end{example}

\subsection{Complete-Data Likelihood Inference}

Even though in general it would be computationally expensive to calculate
the MLE by direct filtering, the maximum complete-data likelihood estimate
(MCLE) is much simpler. A comprehensive analysis of the complete-data
likelihood inference is performed in the following.

Let $N_{t}^{\left( y\right) ,\mathrm{A}}$ and $N_{t}^{\left( y\right) ,%
\mathrm{D}}$ be the counting process of the temporary arrival of size $y$
events and the departure of old size $y$ events during the period $(0,T]$.
Also let%
\begin{equation*}
N_{t}^{\mathrm{type}}\triangleq \sum_{y\in
\mathbb{Z}
\backslash \left\{ 0\right\} }N_{t}^{\left( y\right) ,\mathrm{type}},\text{\
\ \ \ }\mathrm{type}=\mathrm{A},\mathrm{D}.
\end{equation*}

\begin{theorem}
\label{Thm.: MCLE}The complete-data log-likelihood function of the
exponential-trawl process is (ignoring the constant)%
\begin{eqnarray}
l_{\mathcal{C}_{T}}\left( \mathbf{\theta }\right) &=&\sum_{y\in
\mathbb{Z}
\backslash \left\{ 0\right\} }\left( \log \left( \nu \left( y\right) \right)
\left( N_{T}^{\left( y\right) ,\mathrm{A}}+C_{0}^{\left( y\right) }\right)
-\nu \left( y\right) \left( T+\phi ^{-1}\right) \right)
\label{Complete-data log-likelihood with exp trawl} \\
&&+\log \left( \phi \right) \left( N_{T}^{\mathrm{D}}-D_{0}\right) -\phi
\int_{t\in (0,T]}D_{t-}\mathrm{d}t,  \notag
\end{eqnarray}%
so the corresponding MCLE's for the L\'{e}vy measure and the trawl parameter
are%
\begin{eqnarray}
\hat{\nu}_{\mathrm{MCLE}}\left( y\right) &=&\dfrac{N_{T}^{\left( y\right) ,%
\mathrm{A}}+C_{0}^{\left( y\right) }}{T+\hat{\phi}_{\mathrm{MCLE}}^{-1}},\ \
\ \ y\in
\mathbb{Z}
\backslash \left\{ 0\right\} ,  \label{Complete-data MLE} \\
\hat{\phi}_{\mathrm{MCLE}} &=&\frac{\Xi _{T}+\sqrt{\Xi _{T}^{2}+4\dfrac{%
N_{T}^{\mathrm{A}}+N_{T}^{\mathrm{D}}}{T}\int_{t\in (0,T]}D_{t-}\mathrm{d}t}%
}{2\int_{t\in (0,T]}D_{t-}\mathrm{d}t},  \notag \\
\Xi _{T} &\triangleq &N_{T}^{\mathrm{D}}-D_{0}-\dfrac{1}{T}\int_{t\in
(0,T]}D_{t-}\mathrm{d}t.  \notag
\end{eqnarray}%
Furthermore, the MCLE's above are strong consistent: with probability $1$,
as $T\rightarrow \infty $,%
\begin{equation*}
\hat{\phi}_{\mathrm{MCLE}}\rightarrow \phi \text{ and }\hat{\nu}_{\mathrm{%
MCLE}}\left( y\right) \rightarrow \nu \left( y\right) ,\ \ \ \ y\in
\mathbb{Z}
\backslash \left\{ 0\right\} .
\end{equation*}
\end{theorem}

We note that $\widehat{\phi }_{\mathrm{MCLE}}$ depends on $\int_{t\in
(0,T]}D_{t-}$, the total number of possible departures, weighed by time, at
risk during the period $(0,T]$.

\subsection{MLE Calculation based on EM Algorithm\label{Section: EM
Algorithm}}

In this Subsection, we introduce an EM algorithm that is particularly
suitable for exponential-trawl processes, as there are no discretization
errors. The EM algorithm is also computationally efficient. Compared with generic optimization methods like \emph{%
limited-momory BFGS (L-BFGS)}, the updating scheme suggested by EM can
converge to the MLE in a fewer steps and with no error. Clearly, the use of
EM needs some extra computations in each step for backward smoothing, but in
aggregate EM performs much faster than L-BFGS as EM skips those intermediate
filtering calculations during those inactivity periods.

\begin{description}
\item[$E$-Step] The linear form of the complete-data log-likelihood (\ref%
{Complete-data log-likelihood with exp trawl}) allows us to easily take
expectation on it with respect to $\mathbb{P}\left( \cdot |\mathcal{F}%
_{T}\right) $ (under a set of old estimated parameters $\mathbf{\hat{\theta}}%
_{\mathrm{old}}$), which then requires the calculations of the following
quantities using the smoothing distribution $p_{t,T}$:%
\begin{eqnarray}
\mathbb{E}\left( \left. N_{T}^{\left( y\right) ,\mathrm{A}}\right\vert
\mathcal{F}_{T}\right) &=&\sum_{0<t\leq T}\mathbb{P}\left( \left. \Delta
C_{t}^{\left( y\right) }=1\right\vert \mathcal{F}_{T}\right) ,
\label{All the missing data smoothing expectation} \\
\mathbb{E}\left( \left. N_{T}^{\left( y\right) ,\mathrm{D}}\right\vert
\mathcal{F}_{T}\right) &=&\sum_{0<t\leq T}\mathbb{P}\left( \left. \Delta
C_{t}^{\left( y\right) }=-1\right\vert \mathcal{F}_{T}\right) ,  \notag \\
\mathbb{E}\left( \left. C_{0}^{\left( y\right) }\right\vert \mathcal{F}%
_{T}\right) &=&\sum_{\mathbf{j}}j_{y}p_{0,T}\left( \mathbf{j}\right) ,\
\mathbb{E}\left( \left. D_{0}\right\vert \mathcal{F}_{T}\right) =\sum_{%
\mathbf{j}}\left\Vert \mathbf{j}\right\Vert _{1}p_{0,T}\left( \mathbf{j}%
\right) ,  \notag \\
\mathbb{E}\left( \left. D_{t-}\right\vert \mathcal{F}_{T}\right) &=&\sum_{%
\mathbf{j}}\left\Vert \mathbf{j}\right\Vert _{1}p_{t-,T}\left( \mathbf{j}%
\right) ,  \notag
\end{eqnarray}%
where (\ref{Smoothing probability of arrival}) and (\ref{Smoothing
probability of departure}) will be extensively used. Note that $\mathbb{E}%
\left( D_{t-}|\mathcal{F}_{T}\right) $ will be a step function of $t$, so
the calculation of $\int_{t\in (0,T]}\mathbb{E}\left( D_{t-}|\mathcal{F}%
_{T}\right) \mathrm{d}t$ is trivially exact.

\item[$M$-Step] Since the $E$-Step generates a $Q$ function that takes the
same functional form of $\mathbf{\theta }$ as (\ref{Complete-data
log-likelihood with exp trawl}), the solution to $M$-Step takes the same
form as the MCLE in (\ref{Complete-data MLE}), where we just replace each of
the hidden data related terms by their smoothing expectations in (\ref{All
the missing data smoothing expectation}). This can be also viewed as a
representation of plug-in principle for (\ref{Complete-data MLE}), i.e.,
replacing those unknown quantities (e.g. $1_{\left\{ \Delta C_{t}^{\left(
y\right) }=1\right\} }$) by the known ones (e.g. $\mathbb{P}\left( \left.
\Delta C_{t}^{\left( y\right) }=1\right\vert \mathcal{F}_{T}\right) $). We
further use the solution of this $M$-Step for next iteration.
\end{description}

\begin{example}[Continued from Example \protect\ref{Ex.: Skellam Filtering}]

\label{Comparison between EM and L-BFGS} Using the same simulated Skellam
exponential-trawl process path, Table \ref{Tab.: LBFGS vs EM} compares the
MLE derived from (i) the \texttt{L-BFGS-B} procedure in the \texttt{optim}
function of the \texttt{R} language (using the default tolerance settings)
with that from (ii) the EM algorithm (using the same initial parameter
value), which stops until each parameter differs less than a uniform
tolerance $10^{-6}$.
\begin{table}[tbp]
\caption{The MLE calculations on one simulated Skellam exponential-trawl
process using \texttt{L-BFGS-B} procedure in \texttt{R} (with default
settings) and EM algorithm (with uniform tolerance $10^{-6}$ on the
parameter space). The \texttt{R} elapsed time is $137.4$ (sec.) for \texttt{%
L-BFGS-B} and $3.3$ (sec.) for EM, which is about 40 times speed up. Code:
\texttt{EPTprocess\_MLE\_Inference\_Simulation\_LBFGS\_vs\_EM.R}}
\label{Tab.: LBFGS vs EM}\input{LBFGS_vs_EM_table}
\end{table}

As expected, using the EM algorithm gives estimation values that are
very close to the direct optimization of the log-likelihood function (using $%
\delta _{\mathrm{inactivity}}=0.5$). An interesting feature here is that the
MLE found by the EM algorithm has a slightly larger log-likelihood value
(even for $\delta _{\mathrm{inactivity}}=0.01$) than by the \texttt{L-BFGS-B}%
, which might attribute to the numerical insufficiency of the default
optimization tolerance setting of \texttt{R.}

The \texttt{L-BFGS-B} procedure uses $27$ evaluations of the filtering
procedure ($9$ of them for objective function evaluations and $18$ of them
for numerical gradients); as a comparison, the EM algorithm takes $12$
evaluations of the filtering procedure plus $12$ more of the smoothing
procedure. In aggregate, the EM algorithm is over $40$ times faster than the
\texttt{L-BFGS-B} in terms of the computation time.
\end{example}

Starkly different from Example \ref{Ex.: MLE FIltering large vs small delta}%
, the EM algorithm does not require the fine evaluation of the integrals of $%
\lambda _{t-}^{\left( y\right) }$, so not only the filtering procedure in
each iteration of the EM is faster (as it skips the grid calculations of $%
\lambda _{t-}^{\left( y\right) }$ during those inactivity periods) but also
the convergent result of EM will maximize the \emph{numerically errorless}
log-likelihood (as it has nothing to do with $\delta _{\mathrm{inactivity}}$
to conduct EM). As a conclusion, using EM algorithm to search the MLE for
exponential-trawl processes will dominate the direct optimization of
log-likelihood both on the numerical quality and on the computation speed.

\subsection{Likelihood Inference without the Initial Information}

If we consider the complete-data log-likelihood given the information $%
\mathbf{C}_{0}$, i.e. $l_{\mathcal{C}_{T}|\mathbf{C}_{0}}\left( \mathbf{%
\theta }\right) $, then the MCLE's are even simpler:%
\begin{equation*}
\hat{\nu}_{\mathrm{MCLE}}\left( y\right) =\dfrac{N_{T}^{\left( y\right) ,%
\mathrm{A}}}{T},\ \hat{\phi}_{\mathrm{MCLE}}=\frac{N_{T}^{\mathrm{D}}}{%
\int_{t\in (0,T]}D_{t-}\mathrm{d}t}.
\end{equation*}%
Note that these estimates are the most natural frequency estimates providing
that we know the hidden state process $\mathbf{C}_{t}$: $\nu \left( y\right)
$ is estimated by the sample intensity of all the arrivals of size $y$
events, while $\phi ^{-1}$ is estimated by the average lifetime among all
the departures of the temporary events, for the lifetime of any temporary
event is exponentially distributed with mean $1/\phi $.

However, here is a subtle statistical inconsistency if one wants to build an
EM\ algorithm based on $l_{\mathcal{C}_{T}|\mathbf{C}_{0}}\left( \mathbf{%
\theta }\right) $. In practice, all the initial values $C_{0}^{\left(
y\right) }$'s are unknown, so the only way we can work on $l_{\mathcal{C}%
_{T}|\mathbf{C}_{0}}\left( \mathbf{\theta }\right) $ is to treat them as
nuisance parameters. Thus, the EM\ $Q$ function is defined by%
\begin{equation*}
Q\left( \mathbf{\theta }^{\prime },\mathbf{C}_{0}^{\prime }|\mathbf{\theta },%
\mathbf{C}_{0}\right) =\mathbb{E}_{\mathbf{\theta }}\left( \left. l_{%
\mathcal{C}_{T}|\mathbf{C}_{0}^{\prime }}\left( \mathbf{\theta }^{\prime
}\right) \right\vert \mathbf{C}_{0},\mathcal{F}_{T}\right) ,
\end{equation*}%
which not only requires the smoothing scheme based on $\mathbb{P}_{\mathbf{%
\theta }}\left( \cdot |\mathcal{F}_{T},\mathbf{C}_{0}\right) $---not $%
\mathbb{P}_{\mathbf{\theta }}\left( \cdot |\mathcal{F}_{T}\right) $---but
also finally gives us the MLE of the \emph{joint} log-likelihood function $%
l_{\mathcal{F}_{T}}\left( \mathbf{\theta },\mathbf{C}_{0}\right) $---not the
MLE of $l_{\mathcal{F}_{T}}\left( \mathbf{\theta }\right) $ nor of $l_{%
\mathcal{F}_{T}|Y_{0}}\left( \mathbf{\theta }\right) $. On the other hand,
one might also define the EM $Q$ function as%
\begin{equation*}
Q\left( \mathbf{\theta }^{\prime }|\mathbf{\theta }\right) =\mathbb{E}_{%
\mathbf{\theta }}\left( l_{\mathcal{C}_{T}|\mathbf{C}_{0}}\left( \mathbf{%
\theta }^{\prime }\right) |\mathcal{F}_{T}\right) ,
\end{equation*}%
but in this case%
\begin{equation*}
Q\left( \mathbf{\theta }|\mathbf{\theta }\right) =l_{\mathcal{F}%
_{T}|Y_{0}}\left( \mathbf{\theta }\right) -\mathbb{E}_{\mathbf{\theta }%
}\left( l_{\mathbf{C}_{0}|Y_{0}}\left( \mathbf{\theta }\right) |\mathcal{F}%
_{T}\right) \neq l_{\mathcal{F}_{T}|Y_{0}}\left( \mathbf{\theta }\right) ,
\end{equation*}%
which then \emph{breaks} the fundamental monotonicity that guarantees the
availability of EM:%
\begin{equation*}
l_{\mathcal{F}_{T}|Y_{0}}\left( \mathbf{\theta }^{\ast }\right) \geq Q\left(
\mathbf{\theta }^{\ast }|\mathbf{\theta }\right) =\max\limits_{\mathbf{%
\theta }^{\prime }}Q\left( \mathbf{\theta }^{\prime }|\mathbf{\theta }%
\right) \geq Q\left( \mathbf{\theta }|\mathbf{\theta }\right) =l_{\mathcal{F}%
_{T}|Y_{0}}\left( \mathbf{\theta }\right) .
\end{equation*}

Therefore, even though the direct filtering allows the calculations of the
MLE whenever we include the initial information $Y_{0}$ or not (i.e. to
maximize $l_{\mathcal{F}_{T}}\left( \mathbf{\theta }\right) $ or $l_{%
\mathcal{F}_{T}|Y_{0}}\left( \mathbf{\theta }\right) $), a \emph{correct}
EM-based inference will automatically enforce the consideration of $Y_{0}$
(i.e. to maximize $l_{\mathcal{F}_{T}}\left( \mathbf{\theta }\right) $ using
EM). This is a bit different from likelihood inference for marked point
processes, which usually ignores the effect of the initial value $Y_{0}$.
This mild difference will clearly disappear asymptotically as $T\rightarrow
\infty $, but here we still prefer to present a complete likelihood analysis
for trawl processes instead of treating them the same as marked point
processes.

\section{Likelihood Inference for Non-negative Exponential-Trawl Processes
\label{Section: Non-Negative case likelihood}}

In this Section, we focus on exponential-trawl processes that are always
non-negative. Then all the negative movements of this type of processes must
attribute to the departures of the positive events in the trawl, so it is
natural to split up $Y$ into the counting process of size $y$ jumps%
\begin{equation*}
N_{t}^{\left( y\right) }\triangleq \sum_{0<s\leq t}1_{\left\{ \Delta
Y_{s}=y\right\} },\ \ \ \ y\in
\mathbb{Z}
\backslash \left\{ 0\right\} ,
\end{equation*}%
which relates to $C_{t}^{\left( y\right) }\ $via%
\begin{equation}
C_{t}^{\left( y\right) }=C_{0}^{\left( y\right) }+N_{t}^{\left( y\right)
}-N_{t}^{\left( -y\right) }.
\label{C counting process and N counting process}
\end{equation}%
Then, as mentioned in the end of Subsection \ref{Sect: Trawl process
decomposition},%
\begin{equation*}
Y_{t}=\sum_{y=1}^{\infty }yC_{t}^{\left( y\right) }=\sum_{y=1}^{\infty
}yC_{0}^{\left( y\right) }+\sum_{y=1}^{\infty }y(N_{t}^{\left( y\right)
}-N_{t}^{\left( -y\right) }).
\end{equation*}%
Clearly, the path of $Y_{t}$ reveals the path of each of the individual $%
N_{t}^{\left( y\right) }$ for $y\in
\mathbb{Z}
\backslash \left\{ 0\right\} $, so $N_{t}^{\left( y\right) }\in \mathcal{F}%
_{t}$. Thus, the only unknown objects here are $C_{0}^{\left( y\right) }$'s,
for we just see $Y_{0}=\sum_{y=1}^{\infty }yC_{0}^{\left( y\right) }$ and
all the departures resulted from $C_{0}^{\left( y\right) }$'s. If we can
know $C_{0}^{\left( y\right) }$, then we will see the complete path of $%
C_{t}^{\left( y\right) }$ and hence likelihood inference will be
particularly tractable.

\subsection{Partial Likelihood Inference}

We can specialize Corollary \ref{Cor.: log-likelihood for general trawl
processes} using (\ref{exponential trawl non-negative}) and write down the
log-likelihood for the non-negative case (ignoring the constant):%
\begin{eqnarray*}
l_{\mathcal{F}_{T}}\left( \mathbf{\theta }\right) &=&\sum_{y=1}^{\infty
}\left( \log \left( \nu \left( y\right) \right) N_{T}^{\left( y\right) }-\nu
\left( y\right) T\right) \\
&&+\log \left( \phi \right) N_{T}^{\left( -\right) }-\phi \int_{t\in (0,T]}%
\mathbb{E}_{\mathbf{\theta }}\left( \left. C_{t-}^{\left( +\right)
}\right\vert \mathcal{F}_{t-}\right) \mathrm{d}t \\
&&+\sum_{0<t\leq T}\sum_{y=1}^{\infty }\log \mathbb{E}_{\mathbf{\theta }%
}\left( \left. C_{t-}^{(y)}\right\vert \mathcal{F}_{t-}\right) 1_{\left\{
\Delta Y_{t}=-y\right\} }+l_{Y_{0}}\left( \mathbf{\theta }\right) ,
\end{eqnarray*}%
where $N_{T}^{\left( -\right) }\triangleq \sum_{y=1}^{\infty }N_{T}^{\left(
-y\right) }$ and $C_{t-}^{\left( +\right) }\triangleq \sum_{y=1}^{\infty
}C_{t-}^{\left( y\right) }$.

Like the general case we studied in Section \ref{Section: Likelihood
inferences}, there are no analytic expressions available for the filtering
expectations $\mathbb{E}_{\mathbf{\theta }}\left( \left.
C_{t-}^{(y)}\right\vert \mathcal{F}_{t-}\right) $ and the initial likelihood
$l_{Y_{0}}\left( \mathbf{\theta }\right) $, so finding $\mathbf{\hat{\theta}}%
_{\mathrm{MLE}}$ also requires the EM techniques we introduced before.
However, the first part of $l_{\mathcal{F}_{T}}\left( \mathbf{\theta }%
\right) $ that involves $\nu \left( y\right) $'s is particularly
analytically tractable, so this leads us to consider the following maximum
partial likelihood estimate (MPLE) for the L\'{e}vy measure:%
\begin{equation*}
\hat{\nu}_{\mathrm{MPLE}}\left( y\right) =\dfrac{N_{T}^{\left( y\right) }}{T}%
,\ \ \ \ y=1,2,3,...,
\end{equation*}%
which is a non-parametric moment estimate that is apparent from the
non-negative setting.

Even though $\hat{\nu}_{\mathrm{MPLE}}$ is not $\hat{\nu}_{\mathrm{MLE}}$,
it has several advantages. First, it has strong consistency, i.e., with
probability $1$, $\hat{\nu}_{\mathrm{MPLE}}\left( y\right) \rightarrow \nu
\left( y\right) $ as $T\rightarrow \infty $. Second, it is asymptotically
equivalent to the MCLE, because%
\begin{equation*}
\hat{\nu}_{\mathrm{MCLE}}\left( y\right) =\dfrac{N_{T}^{\left( y\right)
}+C_{0}^{\left( y\right) }}{T+\widehat{\phi }_{\mathrm{MCLE}}^{-1}}=\dfrac{%
\dfrac{N_{T}^{\left( y\right) }}{T}+\dfrac{C_{0}^{\left( y\right) }}{T}}{1+%
\dfrac{\widehat{\phi }_{\mathrm{MCLE}}^{-1}}{T}}\approx \hat{\nu}_{\mathrm{%
MPLE}},
\end{equation*}%
where the MCLE of $\mathbf{\theta }$ is simply given from (\ref%
{Complete-data MLE}) but we need to replace those $D_{t-}$ by $%
C_{t-}^{\left( +\right) }$. Third, it allows to estimate each component of
the L\'{e}vy measure separately from themselves and from $\phi $, as given a
long enough path of $Y$, including the initial value $C_{0}^{\left( y\right)
}$ and $\widehat{\phi }_{\mathrm{MCLE}}^{-1}$ has no strong improvement on
the estimation quality of $\hat{\nu}_{\mathrm{MPLE}}$.

Alternatively, a parameterized common intensity function $\nu (y|\eta )$ can
be used, where $\eta $ is some finite dimensional parameter. Then the MPLE
is found by solving%
\begin{equation*}
\hat{\eta}_{\mathrm{MPLE}}\triangleq \limfunc{argmax}\limits_{\eta
}\sum_{y=1}^{\infty }\left( \log \left( \nu \left( y|\eta \right) \right)
N_{T}^{\left( y\right) }-\nu \left( y|\eta \right) T\right)
\end{equation*}%
and letting $\hat{\nu}_{\mathrm{MPLE}}\left( y\right) =\nu \left( y|\hat{\eta%
}_{\mathrm{MPLE}}\right) $.

To infer on the trawl parameter $\phi $, we can simply plug-in the $\hat{\nu}%
_{\mathrm{MPLE}}$ (either parametric or non-parametric) and then do the
filtering procedure to calculate $\mathbb{E}_{\mathbf{\theta }=\left( \hat{%
\nu}_{\mathrm{MPLE}},\phi \right) }\left( \left. C_{t-}^{(y)}\right\vert
\mathcal{F}_{t-}\right) $ for $y=1,2,...$ and $t\in (0,T]$. Combining this
with an (one-dimensional) optimization procedure we can find
\begin{equation*}
\hat{\phi}_{\mathrm{MPLE}}\triangleq \limfunc{argmax}\limits_{\phi }l_{%
\mathcal{F}_{T}}\left( \hat{\nu}_{\mathrm{MPLE}},\phi \right) .
\end{equation*}

\subsection{Estimate the Missing Initial Missing Values}

Except the Poisson case ($Y_{0}=C_{0}^{(1)}$, $C_{0}^{\left( y\right) }=0$
for all $y>1$ and hence in particular $\mathbf{\hat{\theta}}_{\mathrm{MCLE}}=%
\mathbf{\hat{\theta}}_{\mathrm{MLE}}$), every $C_{0}^{\left( y\right) }$'s
are missing, so in principle we need to estimate these initial values in
order to get (an approximation of) $\mathbf{\hat{\theta}}_{\mathrm{MCLE}}$.
Indeed, the EM algorithm also does so through the smoothing expectations%
\begin{equation*}
\mathbb{E}\left( \left. C_{t-}^{\left( y\right) }\right\vert \mathcal{F}%
_{T}\right) =\mathbb{E}\left( \left. C_{0}^{\left( y\right) }\right\vert
\mathcal{F}_{T}\right) +N_{t}^{\left( y\right) }-N_{t}^{\left( -y\right) },
\end{equation*}%
but it just iterates (\ref{Complete-data MLE}) until converges.
Nevertheless, there is another simpler estimation of $C_{0}^{\left( y\right)
}$ thanks to the special non-negative feature.

The following Proposition only relies on the fact that $Y_{t}$ is
non-negative and in fact does not depend on the choice of the trawl.

\begin{proposition}
\label{Prop.: Bound the initial values}Assume that the (general) trawl
process $Y_{t}$ is non-negative. If%
\begin{equation*}
C_{0,T}^{\left( y\right) ,\mathrm{L}}\triangleq \sup_{t\in \left[ 0,T\right]
}\left( N_{t}^{(-y)}-N_{t}^{(y)}\right) ,\ C_{0,T}^{\left( y\right) ,\mathrm{%
U}}\triangleq \left\lfloor \frac{Y_{0}-\sum_{y^{\prime }\neq y}y^{\prime
}C_{0,T}^{\left( y^{\prime }\right) ,\mathrm{L}}}{y}\right\rfloor ,
\end{equation*}%
where $N_{0}^{\left( y\right) }\triangleq 0$ conventionally and $%
\left\lfloor x\right\rfloor $ means the integer part of $x$, then%
\begin{equation*}
C_{0,T}^{\left( y\right) ,\mathrm{U}}\geq C_{0}^{\left( y\right) }\geq
C_{0,T}^{\left( y\right) ,\mathrm{L}}.
\end{equation*}%
Furthermore,%
\begin{equation*}
\lim\limits_{T\rightarrow \infty }C_{0,T}^{\left( y\right) ,\mathrm{U}%
}=\lim\limits_{T\rightarrow \infty }C_{0,T}^{\left( y\right) ,\mathrm{L}%
}=C_{0}^{\left( y\right) }.
\end{equation*}
\end{proposition}

Thus, a straightforward and sharp estimation to $C_{0}^{\left( y\right) }$
can be given by, e.g.,
\begin{equation*}
\hat{C}_{0}^{\left( y\right) }\triangleq \left\lfloor \dfrac{C_{0,T}^{\left(
y\right) ,\mathrm{U}}+C_{0,T}^{\left( y\right) ,\mathrm{L}}}{2}\right\rfloor
,
\end{equation*}%
so use this estimation in (\ref{Complete-data MLE}) will give an estimate of
$\mathbf{\theta }$ that is almost as good as $\mathbf{\hat{\theta}}_{\mathrm{%
MCLE}}$.

\begin{example}
Figure \ref{FIg.: Initial Estimation} illustrates Proposition \ref{Prop.:
Bound the initial values} with a non-negative geometric L\'{e}vy basis, where%
\begin{eqnarray*}
\nu (y|\eta ) &=&\left\Vert \nu \right\Vert \eta \left( 1-\eta \right)
^{y-1},\ \ \ \ y=1,2,..., \\
\left\Vert \nu \right\Vert &=&3,\ \eta =0.5,\ \phi =0.5,\ T=100.
\end{eqnarray*}
The paths of the upper bound $C_{0,t}^{\left( y\right) ,\mathrm{U}}$ and the
lower bound $C_{0,t}^{\left( y\right) ,\mathrm{L}}$ are shown as step
functions of time $t$ in Fig. \ref{FIg.: Initial Estimation}. We can observe
a strong convergent pattern, as all the bounds for different $y$ converge
after $t>15$---the perfect estimations of the initial values $C_{0}^{\left(
y\right) }$'s. Furthermore, as $Y_{0}=C_{0}^{\left( 1\right)
}+2C_{0}^{\left( 2\right) }+4C_{0}^{\left( 4\right) }$ in this case, all the
other $C_{0}^{\left( y\right) }$'s for $y\neq 1,2,4$ must be zero. We then
have discovered all the initial values and can use them to conduct MCLE by (%
\ref{Complete-data MLE}).
\begin{figure}[th]
\centering%
\includegraphics[width=11.7cm]{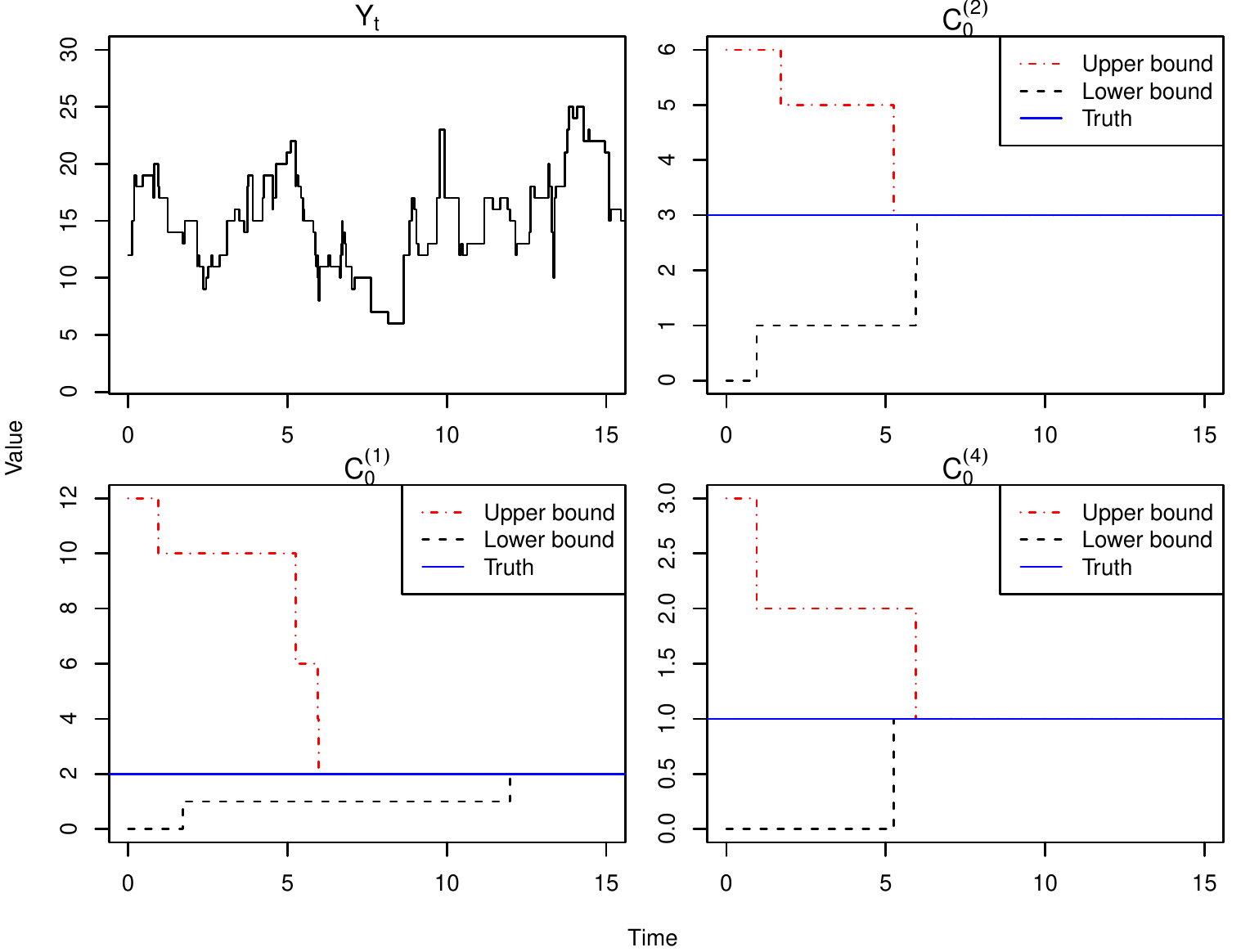}
\caption{\emph{Top left}: A simulated path for the exponential-trawl process
$Y_{t}$ using non-negative geometric L\'{e}vy basis. \emph{Top right}, \emph{%
Bottom left}, \emph{Bottom right}: Paths of $C_{0,t}^{\left( y\right) ,%
\mathrm{U}}$ and $C_{0,t}^{\left( y\right) ,\mathrm{L}}$ along with the true
$C_{0}^{\left( y\right) }$ for $y=1,2,4$. Code: \texttt{EPTprocess%
\_NonNegativeInitialEstimate.R}}
\label{FIg.: Initial Estimation}
\end{figure}
\end{example}

\section{Conclusion\label{Section: Conclude}}

In this Chapter, we studied likelihood-based inference of the trawl
processes by explicitly working on the filtering and smoothing procedures
inherited from this model. It is plausible and practically implementable
under the exponential trawl. We used some simulation examples to justify the
correctness of our procedures.

The major contribution of this Chapter is to provide an easiest beginning
step toward likelihood inference for all of the other more general trawl
processes, which might even allow the inclusion of a non-stationary L\'{e}vy
process component. \cite{ShephardYang(14)} calls it a fleeting price process
and extensively uses it for the study of high frequency financial
econometrics.

The filters for the fleeting price process they proposed will allow an
econometrically interesting decomposition of observed prices into
equilibrium prices and market microstructure noises. More empirical analysis
about these will be addressed in the future work.

\section*{Appendix: Proofs and Derivations}

\addcontentsline{toc}{section}{Appendix}

\subsection{Heuristic Proof of Theorem \protect\ref{Thm.: Point process
Radon-Nykodym Derivative}}

Our heuristic derivation starts from the following prediction decomposition
of the Radon-Nikodym derivative:%
\begin{equation}
\log \left( \dfrac{\mathrm{d}\mathbb{P}}{\mathrm{d}%
\mathbb{Q}
}\right) _{\mathcal{F}_{T}^{X}|X_{0}}=\int_{t\in (0,T]}\log \left( \dfrac{%
\mathrm{d}\mathbb{\mathbb{P}}}{\mathrm{d}%
\mathbb{Q}
}\right) _{X_{t}|\mathcal{F}_{t-}^{X}},
\label{Prediction probability decomposition}
\end{equation}%
where the integral over $t\in (0,T]$ means a continuous sum of the integrand
random variables. Thus,%
\begin{eqnarray*}
\left( \dfrac{\mathrm{d}\mathbb{\mathbb{P}}}{\mathrm{d}%
\mathbb{Q}
}\right) _{X_{t}|\mathcal{F}_{t-}^{X}} &=&\left( \dfrac{\mathrm{d}\mathbb{%
\mathbb{P}}}{\mathrm{d}%
\mathbb{Q}
}\right) _{\Delta X_{t}|\mathcal{F}_{t-}^{X}} \\
&=&\sum_{y\in
\mathbb{Z}
\backslash \left\{ 0\right\} }\dfrac{\mathbb{P}\left( \Delta X_{t}=y|%
\mathcal{F}_{t-}^{X}\right) }{%
\mathbb{Q}
\left( \Delta X_{t}=y|\mathcal{F}_{t-}^{X}\right) }1_{\left\{ \Delta
X_{t}=y\right\} }+\dfrac{\mathbb{P}\left( \Delta X_{t}=0|\mathcal{F}%
_{t-}^{X}\right) }{%
\mathbb{Q}
\left( \Delta X_{t}=0|\mathcal{F}_{t-}^{X}\right) }1_{\left\{ \Delta
X_{t}=0\right\} } \\
&=&\sum_{y\in
\mathbb{Z}
\backslash \left\{ 0\right\} }\dfrac{\lambda _{t-}^{\left( y\right) ,\mathbb{%
P}}\mathrm{d}t}{\lambda _{t-}^{\left( y\right) ,\mathbb{%
\mathbb{Q}
}}\mathrm{d}t}1_{\left\{ \Delta X_{t}=y\right\} }+\dfrac{1-\sum_{y\in
\mathbb{Z}
\backslash \left\{ 0\right\} }\lambda _{t-}^{\left( y\right) ,\mathbb{P}}%
\mathrm{d}t}{1-\sum_{y\in
\mathbb{Z}
\backslash \left\{ 0\right\} }\lambda _{t-}^{\left( y\right) ,%
\mathbb{Q}
}\mathrm{d}t}1_{\left\{ \Delta X_{t}=0\right\} },
\end{eqnarray*}%
where the first equality follows because $X_{t-}$ is known in $\mathcal{F}%
_{t-}^{X}$; the third equality follows from (\ref{Diff. Def. of conditional
intensity}). Therefore, (\ref{Prediction probability decomposition}) can be
rewritten as%
\begin{eqnarray*}
\int_{t\in \left( 0,T\right] }\log \left( \dfrac{\mathrm{d}\mathbb{\mathbb{P}%
}}{\mathrm{d}%
\mathbb{Q}
}\right) _{X_{t}|\mathcal{F}_{t-}^{X}} &=&\sum_{0<t\leq T}\sum_{y\in
\mathbb{Z}
\backslash \left\{ 0\right\} }\log \left( \dfrac{\lambda _{t-}^{\left(
y\right) ,\mathbb{P}}\mathrm{d}t}{\lambda _{t-}^{\left( y\right) ,\mathbb{%
\mathbb{Q}
}}\mathrm{d}t}\right) 1_{\left\{ \Delta X_{t}=y\right\} } \\
&&+\int_{\left\{ t\in \left( 0,T\right] :\Delta X_{t}=0\right\} }\log \left(
\dfrac{1-\sum_{y\in
\mathbb{Z}
\backslash \left\{ 0\right\} }\lambda _{t-}^{\left( y\right) ,\mathbb{P}}%
\mathrm{d}t}{1-\sum_{y\in
\mathbb{Z}
\backslash \left\{ 0\right\} }\lambda _{t-}^{\left( y\right) ,%
\mathbb{Q}
}\mathrm{d}t}\right) \\
&=&\sum_{0<t\leq T}\sum_{y\in
\mathbb{Z}
\backslash \left\{ 0\right\} }\log \left( \dfrac{\lambda _{t-}^{\left(
y\right) ,\mathbb{P}}}{\lambda _{t-}^{\left( y\right) ,\mathbb{%
\mathbb{Q}
}}}\right) 1_{\left\{ \Delta X_{t}=y\right\} } \\
&&-\int_{t\in \left( 0,T\right] }\sum_{y\in
\mathbb{Z}
\backslash \left\{ 0\right\} }\left( \lambda _{t-}^{\left( y\right) ,\mathbb{%
P}}-\lambda _{t-}^{\left( y\right) ,%
\mathbb{Q}
}\right) \mathrm{d}t,
\end{eqnarray*}%
where the second equality follows from $\log \left( 1-x\right) \approx -x$
for small $x$ and $\left\{ t\in \left( 0,T\right] :\Delta X_{t}\neq
0\right\} $ has Lebesgue measure $0$.

\subsection{Heuristic Proof of Theorem \protect\ref{Thm.: Filtering}}

\subsubsection{Update by inactivity}

We want to update $p_{\tau ,\tau }\left( \mathbf{j}\right) $ by
incorporating the information $\mathcal{F}_{\left( \tau ,t\right)
}\triangleq \sigma \left( \left\{ \Delta Y_{s}=0,\ \tau <s<t\right\} \right)
$ using Bayes' Theorem:%
\begin{eqnarray*}
\mathbb{P}\left( \left. \mathbf{C}_{t-}=\mathbf{j}\right\vert \mathcal{F}%
_{t-}\right) &=&\mathbb{P}\left( \left. \mathbf{C}_{\tau }=\mathbf{j}%
\right\vert \mathcal{F}_{t-}\right) =\mathbb{P}\left( \left. \mathbf{C}%
_{\tau }=\mathbf{j}\right\vert \mathcal{F}_{\tau },\mathcal{F}_{\left( \tau
,t\right) }\right) \\
&\propto &\mathbb{P}\left( \mathcal{F}_{\left( \tau ,t\right) }\left\vert
\mathcal{F}_{\tau },\mathbf{C}_{\tau }=\mathbf{j}\right. \right) \mathbb{P}%
\left( \left. \mathbf{C}_{\tau }=\mathbf{j}\right\vert \mathcal{F}_{\tau
}\right) ,
\end{eqnarray*}%
where the first equality holds because there is no activity of $Y_{s}$ for $%
s\in \left( \tau ,t\right) $ and hence the hidden state $\mathbf{C}$ must
stay the same.

Using the prediction decomposition, we have%
\begin{eqnarray*}
\log \mathbb{P}\left( \mathcal{F}_{\left( \tau ,t\right) }\left\vert
\mathcal{F}_{\tau },\mathbf{C}_{\tau }=\mathbf{j}\right. \right)
&=&\int_{s\in \left( \tau ,t\right) }\log \mathbb{P}\left( \Delta Y_{s}=0|%
\mathcal{F}_{\tau },\mathcal{F}_{\left( \tau ,s\right) },\mathbf{C}_{\tau }=%
\mathbf{j}\right) \\
&=&\int_{s\in \left( \tau ,t\right) }\log \left( 1-\sum_{y\in
\mathbb{Z}
\backslash \left\{ 0\right\} }\nu \left( y\right) \mathrm{d}s-\sum_{y\in
\mathbb{Z}
\backslash \left\{ 0\right\} }\phi j_{y}\mathrm{d}s\right) \\
&=&-\sum_{y\in
\mathbb{Z}
\backslash \left\{ 0\right\} }\nu \left( y\right) \left( t-\tau \right)
-\phi \left\Vert \mathbf{j}\right\Vert _{1}\left( t-\tau \right) ,
\end{eqnarray*}%
where the second equality intuitively holds because we know the
instantaneous departure probability of a size $y$ event at time $s$ is $\phi
C_{s-}^{\left( y\right) }\mathrm{d}s$ but $C_{s-}^{\left( y\right) }=C_{\tau
}^{\left( y\right) }=j_{y}$ under $\mathcal{F}_{\left( \tau ,s\right) }$;
the third equality follows from $\log \left( 1-x\right) \approx -x$ for
small $x$. Therefore,%
\begin{equation*}
\mathbb{P}\left( \left. \mathbf{C}_{t-}=\mathbf{j}\right\vert \mathcal{F}%
_{t-}\right) \propto e^{-\phi \left\Vert \mathbf{j}\right\Vert _{1}\left(
t-\tau \right) }\mathbb{P}\left( \left. \mathbf{C}_{\tau }=\mathbf{j}%
\right\vert \mathcal{F}_{\tau }\right) ,
\end{equation*}%
where we throw out the term $\exp \left( -\sum_{y\in
\mathbb{Z}
\backslash \left\{ 0\right\} }\nu \left( y\right) \left( t-\tau \right)
\right) $ because it doesn't depend on $\mathbf{j}$. Normalizing the
equation above leads to the desired result.

\subsubsection{Update by jump}

We want to update $p_{\tau -,\tau -}\left( \mathbf{j}\right) $ by
incorporating the piece of information, $\Delta Y_{\tau }=y$. First note that%
\begin{eqnarray*}
\mathbb{P}\left( \mathbf{C}_{\tau }=\mathbf{j}|\mathcal{F}_{\tau }\right) &=&%
\mathbb{P}\left( \mathbf{C}_{\tau }=\mathbf{j}|\mathcal{F}_{\tau -},\Delta
Y_{\tau }=y\right) \\
&=&\mathbb{P}\left( \left. \mathbf{C}_{\tau }=\mathbf{j},\mathbf{C}_{\tau -}=%
\mathbf{j}-\mathbf{1}^{\left( y\right) }\right\vert \mathcal{F}_{\tau
-},\Delta Y_{\tau }=y\right) \\
&&+\mathbb{P}\left( \left. \mathbf{C}_{\tau }=\mathbf{j},\mathbf{C}_{\tau -}=%
\mathbf{j}+\mathbf{1}^{\left( -y\right) }\right\vert \mathcal{F}_{\tau
-},\Delta Y_{\tau }=y\right) ,
\end{eqnarray*}%
which corresponds to the arrival of a new size $y$ event and the departure
of an old size $-y$ event.

For the first term,%
\begin{eqnarray*}
&&\mathbb{P}\left( \left. \mathbf{C}_{\tau }=\mathbf{j},\mathbf{C}_{\tau -}=%
\mathbf{j}-\mathbf{1}^{\left( y\right) }\right\vert \mathcal{F}_{\tau
-},\Delta Y_{\tau }=y\right) \\
&=&\dfrac{\mathbb{P}\left( \left. \mathbf{C}_{\tau }=\mathbf{j},\mathbf{C}%
_{\tau -}=\mathbf{j}-\mathbf{1}^{\left( y\right) },\Delta Y_{\tau
}=y\right\vert \mathcal{F}_{\tau -}\right) }{\mathbb{P}\left( \left. \Delta
Y_{\tau }=y\right\vert \mathcal{F}_{\tau -}\right) } \\
&=&\dfrac{\mathbb{P}\left( \mathbf{C}_{\tau }=\mathbf{j},\Delta Y_{\tau
}=y\left\vert \mathbf{C}_{\tau -}=\mathbf{j}-\mathbf{1}^{\left( y\right) },%
\mathcal{F}_{\tau -}\right. \right) \mathbb{P}\left( \left. \mathbf{C}_{\tau
-}=\mathbf{j}-\mathbf{1}^{\left( y\right) }\right\vert \mathcal{F}_{\tau
-}\right) }{\mathbb{P}\left( \left. \Delta Y_{\tau }=y\right\vert \mathcal{F}%
_{\tau -}\right) } \\
&=&\dfrac{\mathbb{P}\left( \Delta \mathbf{C}_{\tau }=\mathbf{1}^{\left(
y\right) }\left\vert \mathbf{C}_{\tau -}=\mathbf{j}-\mathbf{1}^{\left(
y\right) },\mathcal{F}_{\tau -}\right. \right) \mathbb{P}\left( \left.
\mathbf{C}_{\tau -}=\mathbf{j}-\mathbf{1}^{\left( y\right) }\right\vert
\mathcal{F}_{\tau -}\right) }{\mathbb{P}\left( \left. \Delta Y_{\tau
}=y\right\vert \mathcal{F}_{\tau -}\right) } \\
&=&\dfrac{\nu \left( y\right) }{\lambda _{\tau -}^{\left( y\right) }}\mathbb{%
P}\left( \left. \mathbf{C}_{\tau -}=\mathbf{j}-\mathbf{1}^{\left( y\right)
}\right\vert \mathcal{F}_{\tau -}\right) ,
\end{eqnarray*}%
where the fourth equality follows from (\ref{Transition probability of
vector C}) (using $\mathcal{C}_{\tau -}\supseteq \mathcal{F}_{\tau -}$) and (%
\ref{Diff. Def. of conditional intensity}).

Using similar arguments, the second term is%
\begin{equation*}
\mathbb{P}\left( \left. \mathbf{C}_{\tau }=\mathbf{j},\mathbf{C}_{\tau -}=%
\mathbf{j}+\mathbf{1}^{\left( -y\right) }\right\vert \mathcal{F}_{\tau
-},\Delta Y_{\tau }=y\right) =\dfrac{\phi \left( j_{-y}+1\right) }{\lambda
_{\tau -}^{\left( y\right) }}\mathbb{P}\left( \left. \mathbf{C}_{\tau -}=%
\mathbf{j}+\mathbf{1}^{\left( -y\right) }\right\vert \mathcal{F}_{\tau
-}\right) .
\end{equation*}%
Combining all of these gives us the required result.

\subsection{Heuristic Proof of Theorem \protect\ref{Thm.: Smoothing}}

The case of updating smoothing distribution $p_{\tau -,T}\left( \mathbf{j}%
\right) $ due to inactivity is trivial because the hidden configuration $%
\mathbf{C}$ must stay unchanged because of the inactivity during the time
period $[t,\tau )$.

\subsubsection{Update by jump}

We now consider the case of (backward) updating the smoothing distribution $%
p_{\tau ,T}\left( \mathbf{j}\right) $ due to the jump $\Delta Y_{\tau }=y$.
Then%
\begin{eqnarray*}
\mathbb{P}\left( \mathbf{C}_{\tau -}=\mathbf{j}|\mathcal{F}_{T}\right) &=&%
\mathbb{P}\left( \left. \mathbf{C}_{\tau -}=\mathbf{j},\mathbf{C}_{\tau }=%
\mathbf{j}+\mathbf{1}^{\left( y\right) }\right\vert \mathcal{F}_{T}\right) +%
\mathbb{P}\left( \left. \mathbf{C}_{\tau -}=\mathbf{j},\mathbf{C}_{\tau }=%
\mathbf{j}-\mathbf{1}^{\left( -y\right) }\right\vert \mathcal{F}_{T}\right)
\\
&=&\mathbb{P}\left( \mathbf{C}_{\tau -}=\mathbf{j}\left\vert \mathcal{F}_{T},%
\mathbf{C}_{\tau }=\mathbf{j}+\mathbf{1}^{\left( y\right) }\right. \right)
\mathbb{P}\left( \left. \mathbf{C}_{\tau }=\mathbf{j}+\mathbf{1}^{\left(
y\right) }\right\vert \mathcal{F}_{T}\right) \\
&&+\mathbb{P}\left( \mathbf{C}_{\tau -}=\mathbf{j}\left\vert \mathcal{F}_{T},%
\mathbf{C}_{\tau }=\mathbf{j}-\mathbf{1}^{\left( -y\right) }\right. \right)
\mathbb{P}\left( \left. \mathbf{C}_{\tau }=\mathbf{j}-\mathbf{1}^{\left(
-y\right) }\right\vert \mathcal{F}_{T}\right) .
\end{eqnarray*}

Note that%
\begin{eqnarray}
\mathbb{P}\left( \mathbf{C}_{\tau -}=\mathbf{j}\left\vert \mathcal{F}_{T},%
\mathbf{C}_{\tau }=\mathbf{k}\right. \right) &=&\mathbb{P}\left( \mathbf{C}%
_{\tau -}=\mathbf{j}\left\vert \mathcal{F}_{\tau },\mathbf{C}_{\tau }=%
\mathbf{k}\right. \right)  \label{The smoothing subtleties} \\
&=&\dfrac{\mathbb{P}\left( \mathbf{C}_{\tau }=\mathbf{k}|\mathbf{C}_{\tau -}=%
\mathbf{j},\mathcal{F}_{\tau }\right) \mathbb{P}\left( \mathbf{C}_{\tau -}=%
\mathbf{j}|\mathcal{F}_{\tau }\right) }{\mathbb{P}\left( \mathbf{C}_{\tau }=%
\mathbf{k}|\mathcal{F}_{\tau }\right) }  \notag \\
&=&\dfrac{%
\begin{array}{c}
\mathbb{P}\left( \mathbf{C}_{\tau }=\mathbf{k}|\mathbf{C}_{\tau -}=\mathbf{j}%
,\mathcal{F}_{\tau }\right) \mathbb{P}\left( \Delta Y_{\tau }=y|\mathbf{C}%
_{\tau -}=\mathbf{j},\mathcal{F}_{\tau -}\right) \\
\times \mathbb{P}\left( \mathbf{C}_{\tau -}=\mathbf{j}|\mathcal{F}_{\tau
-}\right)%
\end{array}%
}{\mathbb{P}\left( \mathbf{C}_{\tau }=\mathbf{k}|\mathcal{F}_{\tau }\right)
\mathbb{P}\left( \Delta Y_{\tau }=y|\mathcal{F}_{\tau -}\right) }  \notag \\
&=&\dfrac{\mathbb{P}\left( \mathbf{C}_{\tau }=\mathbf{k},\Delta Y_{\tau }=y|%
\mathbf{C}_{\tau -}=\mathbf{j},\mathcal{F}_{\tau -}\right) }{\lambda _{\tau
-}^{\left( y\right) }\mathrm{d}t}\dfrac{\mathbb{P}\left( \mathbf{C}_{\tau -}=%
\mathbf{j}|\mathcal{F}_{\tau -}\right) }{\mathbb{P}\left( \mathbf{C}_{\tau }=%
\mathbf{k}|\mathcal{F}_{\tau }\right) },  \notag
\end{eqnarray}%
where the first equality holds due to the Markov property of $\mathbf{C}_{t}$%
, a heuristic derivation is given later; the second and third equalities
follow from the Bayes' Theorem. Since%
\begin{eqnarray*}
\mathbb{P}\left( \left. \mathbf{C}_{\tau }=\mathbf{j}+\mathbf{1}^{\left(
y\right) },\Delta Y_{\tau }=y\right\vert \mathbf{C}_{\tau -}=\mathbf{j},%
\mathcal{F}_{\tau -}\right) &=&\mathbb{P}\left( \left. \Delta \mathbf{C}%
_{\tau }=\mathbf{1}^{\left( y\right) }\right\vert \mathbf{C}_{\tau -}=%
\mathbf{j},\mathcal{F}_{\tau -}\right) \\
&=&\nu \left( y\right) \mathrm{d}t, \\
\mathbb{P}\left( \left. \mathbf{C}_{\tau }=\mathbf{j}-\mathbf{1}^{\left(
-y\right) },\Delta Y_{\tau }=y\right\vert \mathbf{C}_{\tau -}=\mathbf{j},%
\mathcal{F}_{\tau -}\right) &=&\mathbb{P}\left( \left. \Delta \mathbf{C}%
_{\tau }=-\mathbf{1}^{\left( -y\right) }\right\vert \mathbf{C}_{\tau -}=%
\mathbf{j},\mathcal{F}_{\tau -}\right) \\
&=&\phi j_{-y}\mathrm{d}t,
\end{eqnarray*}%
combining all of these gives us the required result.

\subsubsection{Derivation of (\protect\ref{The smoothing subtleties})}

Let $\mathcal{F}_{(\tau ,T]}\triangleq \sigma \left( \left\{ Y_{t}\right\}
_{\tau <t\leq T}\right) $ and $\mathcal{C}_{(\tau ,T]}\triangleq \sigma
\left( \left\{ \mathbf{C}_{t}\right\} _{\tau <t\leq T}\right) $. Note that
heuristically the Bayes' Theorem implies%
\begin{eqnarray*}
\mathbb{P}\left( \mathbf{C}_{\tau -}=\mathbf{j}\left\vert \mathcal{F}_{T},%
\mathbf{C}_{\tau }=\mathbf{k}\right. \right) &=&\mathbb{P}\left( \mathbf{C}%
_{\tau -}=\mathbf{j}\left\vert \mathcal{F}_{\tau },\mathcal{F}_{(\tau ,T]},%
\mathbf{C}_{\tau }=\mathbf{k}\right. \right) \\
&=&\dfrac{\left( \dfrac{\mathrm{d}\mathbb{P}}{\mathrm{d}%
\mathbb{Q}
}\right) _{\mathcal{F}_{(\tau ,T]}\left\vert \mathcal{F}_{\tau },\mathbf{C}%
_{\tau }=\mathbf{k},\mathbf{C}_{\tau -}=\mathbf{j}\right. }}{\left( \dfrac{%
\mathrm{d}\mathbb{P}}{\mathrm{d}%
\mathbb{Q}
}\right) _{\mathcal{F}_{(\tau ,T]}\left\vert \mathcal{F}_{\tau },\mathbf{C}%
_{\tau }=\mathbf{k}\right. }}\mathbb{P}\left( \mathbf{C}_{\tau -}=\mathbf{j}%
\left\vert \mathcal{F}_{\tau },\mathbf{C}_{\tau }=\mathbf{k}\right. \right) .
\end{eqnarray*}%
Since $\mathcal{F}_{(\tau ,T]}\subseteq \mathcal{C}_{(\tau ,T]}$ (each $%
Y_{t}=\sum_{y\in
\mathbb{Z}
\backslash \left\{ 0\right\} }C_{t}^{\left( y\right) }$), the Markov
property of $\mathbf{C}_{t}$ implies%
\begin{equation*}
\left( \dfrac{\mathrm{d}\mathbb{P}}{\mathrm{d}%
\mathbb{Q}
}\right) _{\mathcal{F}_{(\tau ,T]}\left\vert \mathcal{F}_{\tau },\mathbf{C}%
_{\tau }=\mathbf{k},\mathbf{C}_{\tau -}=\mathbf{j}\right. }=\left( \dfrac{%
\mathrm{d}\mathbb{P}}{\mathrm{d}%
\mathbb{Q}
}\right) _{\mathcal{F}_{(\tau ,T]}\left\vert \mathcal{F}_{\tau },\mathbf{C}%
_{\tau }=\mathbf{k}\right. },
\end{equation*}%
because given the current information $\mathbf{C}_{\tau }$ the information
in the past $\mathbf{C}_{\tau -}$ is irrelevant. This then proves that%
\begin{equation*}
\mathbb{P}\left( \mathbf{C}_{\tau -}=\mathbf{j}\left\vert \mathcal{F}_{T},%
\mathbf{C}_{\tau }=\mathbf{k}\right. \right) =\mathbb{P}\left( \mathbf{C}%
_{\tau -}=\mathbf{j}\left\vert \mathcal{F}_{\tau },\mathbf{C}_{\tau }=%
\mathbf{k}\right. \right) .
\end{equation*}

\subsection{Proof of Theorem \protect\ref{Thm.: MCLE}}

Since each $C_{t}^{\left( y\right) }$ is independent for different $y$, the
complete-data log-likelihood can be written as%
\begin{equation*}
l_{\mathcal{C}_{T}}\left( \mathbf{\theta }\right) =\sum_{y\in
\mathbb{Z}
\backslash \left\{ 0\right\} }l_{\mathcal{C}_{T}^{\left( y\right)
}|C_{0}^{\left( y\right) }}\left( \mathbf{\theta }\right) +\sum_{y\in
\mathbb{Z}
\backslash \left\{ 0\right\} }l_{C_{0}^{\left( y\right) }}\left( \mathbf{%
\theta }\right) ,
\end{equation*}%
where we recall that $\mathcal{C}_{t}^{\left( y\right) }$ is the natural
filtration generated by $C_{t}^{\left( y\right) }$,%
\begin{eqnarray*}
l_{\mathcal{C}_{T}^{\left( y\right) }|C_{0}^{\left( y\right) }}\left(
\mathbf{\theta }\right) &=&\sum_{0<t\leq T}\left( \log \left( \nu \left(
y\right) \right) 1_{\left\{ \Delta C_{t}^{\left( y\right) }=1\right\} }+\log
\left( \phi C_{t-}^{\left( y\right) }\right) 1_{\left\{ \Delta C_{t}^{\left(
y\right) }=-1\right\} }\right) \\
&&-\int_{t\in (0,T]}\left( \nu \left( y\right) +\phi C_{t-}^{\left( y\right)
}\right) \mathrm{d}t \\
&=&\log \left( \nu \left( y\right) \right) N_{T}^{\left( y\right) ,\mathrm{A}%
}-\nu \left( y\right) T+\log \left( \phi \right) N_{T}^{\left( y\right) ,%
\mathrm{D}}-\phi \int_{t\in (0,T]}C_{t-}^{\left( y\right) }\mathrm{d}t,
\end{eqnarray*}%
where the first equality follows directly from Theorem \ref{Thm.: Point
process Radon-Nykodym Derivative} (ignoring the constant), and%
\begin{equation*}
l_{C_{0}^{\left( y\right) }}\left( \mathbf{\theta }\right) =C_{0}^{\left(
y\right) }\left( \log \nu \left( y\right) -\log \phi \right) -\dfrac{\nu
\left( y\right) }{\phi }
\end{equation*}%
because of $C_{0}^{\left( y\right) }\backsim \mathrm{Poisson}\left( \nu
\left( y\right) /\phi \right) $. Thus, collecting terms will give us the
required result (\ref{Complete-data log-likelihood with exp trawl}). The
derivations of the MCLE are elementary.

Let%
\begin{equation*}
\left\Vert \nu \right\Vert \triangleq \int \nu \left( \mathrm{d}y\right)
=\sum_{y=1}^{\infty }\nu \left( y\right) .
\end{equation*}%
The ergodicity of $D_{t-}$ implies that as $T\rightarrow \infty $%
\begin{equation*}
\dfrac{1}{T}\int_{t\in (0,T]}D_{t-}\mathrm{d}t\rightarrow \mathbb{E}\left(
D_{t-}\right) =\dfrac{\left\Vert \nu \right\Vert }{\phi }.
\end{equation*}%
Since $\dfrac{N_{T}^{\mathrm{D}}}{T}\approx \dfrac{N_{T}^{\mathrm{A}}}{T}%
\rightarrow \left\Vert \nu \right\Vert $, we have%
\begin{equation*}
\dfrac{\Xi _{T}}{T}=\dfrac{N_{T}^{\mathrm{D}}}{T}-\dfrac{D_{0}+T^{-1}\int_{t%
\in (0,T]}D_{t-}\mathrm{d}t}{T}\rightarrow \left\Vert \nu \right\Vert \text{%
, too.}
\end{equation*}%
Thus,%
\begin{eqnarray*}
\hat{\phi}_{\mathrm{MCLE}} &=&\frac{\dfrac{\Xi _{T}}{T}+\sqrt{\left( \dfrac{%
\Xi _{T}}{T}\right) ^{2}+4T^{-1}\dfrac{N_{T}^{\mathrm{A}}+N_{T}^{\mathrm{D}}%
}{T}T^{-1}\int_{t\in (0,T]}D_{t-}\mathrm{d}t}}{2T^{-1}\int_{t\in (0,T]}D_{t-}%
\mathrm{d}t} \\
&\rightarrow &\frac{\left\Vert \nu \right\Vert +\sqrt{\left\Vert \nu
\right\Vert ^{2}+0}}{2\dfrac{\left\Vert \nu \right\Vert }{\phi }}=\phi .
\end{eqnarray*}%
Finally, for any $y\in
\mathbb{Z}
\backslash \left\{ 0\right\} $, $\dfrac{N_{T}^{\left( y\right) }}{T}%
\rightarrow \nu \left( y\right) $ and $\hat{\phi}_{\mathrm{MCLE}%
}^{-1}\rightarrow \phi ^{-1}<\infty $, so we easily have
\begin{equation*}
\hat{\nu}_{\mathrm{MCLE}}\left( y\right) =\dfrac{\dfrac{N_{T}^{\left(
y\right) }}{T}+\dfrac{C_{0}^{\left( y\right) }}{T}}{1+\dfrac{\hat{\phi}_{%
\mathrm{MCLE}}^{-1}}{T}}\rightarrow \nu \left( y\right) \text{ as well.}
\end{equation*}

\subsection{Proof of Proposition \protect\ref{Prop.: Bound the initial
values}}

As $C_{t}^{(y)}\geq 0$, (\ref{C counting process and N counting process})
implies that%
\begin{equation*}
C_{0}^{\left( y\right) }\geq C_{0,T}^{\left( y\right) ,\mathrm{L}%
}=\sup_{t\in \left[ 0,T\right] }\left( N_{t}^{(-y)}-N_{t}^{(y)}\right) ,\ \
\ \ y=1,2,...,
\end{equation*}%
where we set $N_{0}^{\left( y\right) }\triangleq 0$ conventionally. Now%
\begin{equation*}
C_{0}^{\left( y\right) }=\frac{Y_{0}-\sum_{y^{\prime }\neq y}y^{\prime
}C_{0}^{(y^{\prime })}}{y}\leq \left\lfloor \frac{Y_{0}-\sum_{y^{\prime
}\neq y}y^{\prime }C_{0,T}^{\left( y^{\prime }\right) ,\mathrm{L}}}{y}%
\right\rfloor =C_{0,T}^{\left( y\right) ,\mathrm{U}},
\end{equation*}%
so we have%
\begin{equation*}
C_{0,T}^{\left( y\right) ,\mathrm{U}}\geq C_{0}^{\left( y\right) }\geq
C_{0,T}^{\left( y\right) ,\mathrm{L}}.
\end{equation*}

Let $N_{t}^{\left( -y\right) ,\ast }$ be the counting process of $-y$ jumps
resulted from the departures of those initial events of size $y$ that
constitute $C_{0}^{\left( y\right) }$. Let $\tau $ be the time when $%
N^{\left( -y\right) ,\ast }$ achieve $C_{0}^{\left( y\right) }$. Then we have%
\begin{eqnarray*}
C_{0,T}^{\left( y\right) ,\mathrm{L}} &=&C_{0,\tau }^{\left( y\right) ,%
\mathrm{L}}\vee \sup\limits_{t\in (\tau ,T]}\left( N_{t}^{\left( -y\right)
,\ast }-\left( N_{t}^{\left( y\right) }-\left( N_{t}^{\left( -y\right)
}-N_{t}^{\left( -y\right) ,\ast }\right) \right) \right) \\
&=&C_{0,\tau }^{\left( y\right) ,\mathrm{L}}\vee \left( C_{0}^{\left(
y\right) }-\inf\limits_{t\in (\tau ,T]}\left( N_{t}^{\left( y\right)
}-\left( N_{t}^{\left( -y\right) }-N_{t}^{\left( -y\right) ,\ast }\right)
\right) \right) .
\end{eqnarray*}%
Observe that $N_{t}^{\left( y\right) }-\left( N_{t}^{\left( -y\right)
}-N_{t}^{\left( -y\right) ,\ast }\right) $ is a \textrm{M}/$G$/$\infty $
queue initiated at state $0$, so by the ergodicity we must have with
probability $1$%
\begin{equation*}
\lim\limits_{T\rightarrow \infty }\inf\limits_{t\in (\tau ,T]}\left(
N_{t}^{\left( y\right) }-\left( N_{t}^{\left( -y\right) }-N_{t}^{\left(
-y\right) ,\ast }\right) \right) =0.
\end{equation*}%
This then shows that actually%
\begin{equation*}
\lim\limits_{T\rightarrow \infty }C_{0,T}^{\left( y\right) ,\mathrm{L}%
}=C_{0,\tau }^{\left( y\right) ,\mathrm{L}}\vee C_{0}^{\left( y\right)
}=C_{0}^{\left( y\right) },
\end{equation*}%
where the last equality follows because $C_{0,\tau }^{\left( y\right) ,%
\mathrm{L}}\leq C_{0}^{\left( y\right) }$. Correspondingly,%
\begin{equation*}
\lim\limits_{T\rightarrow \infty }C_{0,T}^{\left( y\right) ,\mathrm{U}%
}=\left\lfloor \frac{Y_{0}-\sum_{y^{\prime }\neq y}y^{\prime }C_{0}^{\left(
y\right) }}{y}\right\rfloor =C_{0}^{\left( y\right) }\text{.}
\end{equation*}

\input{referenc}

\end{document}

%% file: preamble.tex
\usepackage{mathptmx}       
\usepackage{helvet}         
\usepackage{courier}        
\usepackage{type1cm}        
\usepackage{makeidx}         
\usepackage{graphicx}        
\usepackage{multicol}        
\usepackage[bottom]{footmisc}
\usepackage{epstopdf}
\DeclareGraphicsExtensions{.pdf}

\makeindex             
\usepackage{url} 
\graphicspath{{../../graphs/}} 

%% file: preintro.tex
\title*{Likelihood Inference for Exponential-Trawl Processes}
\author{Neil Shephard and Justin J. Yang}
\institute{Neil Shephard \at Department of Economics, Harvard University\\
1802 Cambridge Street, 02138, Cambridge, MA, USA\\
\email{shephard@fas.harvard.edu}
\and Justin J. Yang \at Department of Statistics, Harvard University\\
1 Oxford Street, 02138, Cambridge, MA, USA\\
\email{juchenjustinyang@fas.harvard.edu}}
%
%
\maketitle

\abstract*{Integer-valued trawl processes are a class of serially correlated, stationary and infinitely divisible processes that Ole E. Barndorff-Nielsen has been working on in recent years. In this Chapter, we provide the first analysis of likelihood inference for trawl processes by focusing on the so-called exponential-trawl process, which is also a continuous time hidden Markov process with countable state space. The core ideas include prediction decomposition, filtering and smoothing, complete-data analysis and EM algorithm. These can be easily scaled up to adapt to more general trawl processes but with increasing computation efforts.}

\abstract{Integer-valued trawl processes are a class of serially correlated, stationary and infinitely divisible processes that Ole E. Barndorff-Nielsen has been working on in recent years. In this Chapter, we provide the first analysis of likelihood inference for trawl processes by focusing on the so-called exponential-trawl process, which is also a continuous time hidden Markov process with countable state space. The core ideas include prediction decomposition, filtering and smoothing, complete-data analysis and EM algorithm. These can be easily scaled up to adapt to more general trawl processes but with increasing computation efforts.}

%% file: LBFGS_vs_EM_table.tex
\begin{tabular*}{11.7cm}{@{\extracolsep{\fill} }ccccccc}
\hline\noalign{\smallskip}
                  & \multicolumn{3}{c}{Parameter} & & \multicolumn{2}{c}{Log-likelihood}\\
\cline{2-4} \cline{6-7}
Estimation & $\nu ^{+}$ & $\nu ^{-}$ & $\phi $ & & $\delta _{\mathrm{inactivity}}=0.5$ & $\delta _{\mathrm{inactivity}}=0.01$ \\
\noalign{\smallskip}\svhline\noalign{\smallskip}
Truth             & $0.01260$ & $0.01111$ & $0.03402$ & & $-15,974.98$ & $-15,974.9543$\\
\texttt{L-BFGS-B} & $0.01201$ & $0.01128$ & $0.03362$ & & $-15,973.92$ & $-15,973.8915$\\
EM              & $0.01199$ & $0.01126$ & $0.03354$ & & $-15,973.91$ & $-15,973.8881$\\
\noalign{\smallskip}\hline\noalign{\smallskip}
\end{tabular*} 

%% file: referenc.tex
%
%
 \bibliographystyle{spmpsci}
 \bibliography{../Main-AddOn,../BN_80BD,../neil}

\biblstarthook{References may be \textit{cited} in the text either by number (preferred) or by author/year.\footnote{Make sure that all references from the list are cited in the text. Those not cited should be moved to a separate \textit{Further Reading} section or chapter.} The reference list should ideally be \textit{sorted} in alphabetical order -- even if reference numbers are used for the their citation in the text. If there are several works by the same author, the following order should be used:
\begin{enumerate}
\item all works by the author alone, ordered chronologically by year of publication
\item all works by the author with a coauthor, ordered alphabetically by coauthor
\item all works by the author with several coauthors, ordered chronologically by year of publication.
\end{enumerate}
The \textit{styling} of references\footnote{Always use the standard abbreviation of a journal's name according to the ISSN \textit{List of Title Word Abbreviations}, see \url{http://www.issn.org/en/node/344}} depends on the subject of your book:
\begin{itemize}
\item The \textit{two} recommended styles for references in books on \textit{mathematical, physical, statistical and computer sciences} are depicted in ~\cite{science-contrib, science-online, science-mono, science-journal, science-DOI} and ~\cite{phys-online, phys-mono, phys-journal, phys-DOI, phys-contrib}.
\item Examples of the most commonly used reference style in books on \textit{Psychology, Social Sciences} are~\cite{psysoc-mono, psysoc-online,psysoc-journal, psysoc-contrib, psysoc-DOI}.
\item Examples for references in books on \textit{Humanities, Linguistics, Philosophy} are~\cite{humlinphil-journal, humlinphil-contrib, humlinphil-mono, humlinphil-online, humlinphil-DOI}.
\item Examples of the basic Springer style used in publications on a wide range of subjects such as \textit{Computer Science, Economics, Engineering, Geosciences, Life Sciences, Medicine, Biomedicine} are ~\cite{basic-contrib, basic-online, basic-journal, basic-DOI, basic-mono}.
\end{itemize}
} 

%% file: Fleeting_Prices_BN_v1.5.bbl
\begin{thebibliography}{10}
\providecommand{\url}[1]{{#1}}
\providecommand{\urlprefix}{URL }
\expandafter\ifx\csname urlstyle\endcsname\relax
  \providecommand{\doi}[1]{DOI~\discretionary{}{}{}#1}\else
  \providecommand{\doi}{DOI~\discretionary{}{}{}\begingroup
  \urlstyle{rm}\Url}\fi

\bibitem{Asmussen(03)}
Asmussen, S.: Applied Probability and Queues.
\newblock Springer, New York
\newblock  (2003)

\bibitem{BarndorffNielsen(11)}
Barndorff-Nielsen, O.E.: Stationary infinitely divisible processes.
\newblock Braz. J. Probab. Stat. \textbf{25}, 294--322
\newblock  (2011)

\bibitem{BarndorffNielsenBenthVeraart(11)}
Barndorff-Nielsen, O.E., Benth, F.E., Veraart, A.E.D.: Ambit processes and
  stochastic partial differential equations.
\newblock In: G.~Di~Nunno, B.~{\O}ksendal (eds.) Advanced Mathematical Methods
  for Finance, pp. 35--74. Springer, Berlin Heidelberg
\newblock  (2011)

\bibitem{BarndorffNielsenBenthVeraart(12)}
Barndorff-Nielsen, O.E., Benth, F.E., Veraart, A.E.D.: {Recent advances in
  ambit stochastics with a view towards tempo-spatial stochastic
  volatility/intermittency}.
\newblock ArXiv e-prints
\newblock  (2012).
\newblock Unpublished paper, Department of Mathematics, Imperial College London

\bibitem{BarndorffNielsenLundeShephardVeraart(14)}
Barndorff-Nielsen, O.E., Lunde, A., Shephard, N., Veraart, A.E.D.:
  Integer-valued trawl processes: A class of stationary infinitely divisible
  processes.
\newblock Scandinavian Journal of Statistics \textbf{41}, 693--724
\newblock  (2014)

\bibitem{BarndorffNielsenPollardShephard(12)}
Barndorff-Nielsen, O.E., Pollard, D.G., Shephard, N.: Integer-valued
  \uppercase{L}\'evy processes and low latency financial econometrics.
\newblock Quantitative Finance \textbf{12}, 587--605
\newblock  (2012)

\bibitem{BarndorffNielsenSchmiegel(07)}
Barndorff-Nielsen, O.E., Schmiegel, J.: Ambit processes; with applications to
  turbulence and tumour growth.
\newblock In: F.E. Benth, G.~Di~Nunno, T.~Lindstr{\o}m, B.~{\O}ksendal,
  T.~Zhang (eds.) Stochastic Analysis and Applications, pp. 93--124. Springer,
  Berlin Heidelberg
\newblock  (2007)

\bibitem{Bartlett(78)}
Bartlett, M.S.: An Introduction to Stochastic Processes, with Special Reference
  to Methods and Applications, 3 edn.
\newblock Cambridge University Press, Cambridge
\newblock  (1978)

\bibitem{CameronTrivedi(98)}
Cameron, C.A., Trivedi, P.K.: Regression Analysis of Count Data.
\newblock Cambridge University Press, Cambridge
\newblock  (1998)

\bibitem{CuiLund(09)}
Cui, Y., Lund, R.: A new look at time series of counts.
\newblock Biometrika \textbf{96}, 781--792
\newblock  (2009)

\bibitem{DaleyVere-Jones(08_Ch14)}
Daley, D., Vere-Jones, D.: Evolutionary processes and predictability.
\newblock In: An Introduction to the Theory of Point Processes, Probability and
  Its Applications, pp. 355--456. Springer, New York
\newblock  (2008)

\bibitem{DavisWu(09)}
Davis, R.A., Wu, R.: A negative binomial model for time series of counts.
\newblock Biometrika \textbf{96}, 735--749
\newblock  (2009)

\bibitem{KedemFokianos(02)}
Fokianos, K., Kedem, B.: Regression theory for categorical time series.
\newblock Statistical Science \textbf{18}, 357--376
\newblock  (2003)

\bibitem{JacobsLewis(78)}
Jacobs, P.A., Lewis, P.A.W.: Discrete time series generated by mixtures. {I}:
  Correlational and runs properties.
\newblock Journal of the Royal Statistical Society. Series B (Methodological)
  \textbf{40}, 94--105
\newblock  (1978)

\bibitem{JungTremayne(11)}
Jung, R., Tremayne, A.: Useful models for time series of counts or simply wrong
  ones?
\newblock AStA Advances in Statistical Analysis \textbf{95}, 59--91
\newblock  (2011)

\bibitem{Lindley(56)}
Lindley, D.V.: The estimation of velocity distributions from counts.
\newblock In: Proceedings of the Internationl Congress of Mathematicians,
  vol.~3, pp. 427--444. North-Holland, Amsterdam
\newblock  (1956)

\bibitem{McKenzie(03)}
McKenzie, D.J.: Measuring inequality with asset indicators.
\newblock Journal of Population Economics \textbf{18}, 229--260
\newblock  (2005)

\bibitem{Reynolds(68)}
Reynolds, J.F.: On the autocorrelation and spectral functions of queues.
\newblock Journal of Applied Probability \textbf{5}, 467--475
\newblock  (1968)

\bibitem{Rudemo(73)}
Rudemo, M.: State estimation for partially observed {Markov} chains.
\newblock Journal of Mathematical Analysis and Applications \textbf{44},
  581--611
\newblock  (1973)

\bibitem{Rudemo(75)}
Rudemo, M.: Prediction and smoothing for partially observed {Markov} chains.
\newblock Journal of Mathematical Analysis and Applications \textbf{49}, 1--23
\newblock  (1975)

\bibitem{ShephardYang(14)}
Shephard, N., Yang, J.J.: Continuous time analysis of fleeting discrete price
  moves.
\newblock ArXiv e-prints
\newblock  (2014).
\newblock Unpublished paper, Department of Statistics, Harvard Unviersity

\bibitem{SurgailisRosinskiMandrekarCambanis(93)}
Surgailis, D., Rosinski, J., Mandrekar, V., Cambanis, S.: Stable mixed moving
  averages.
\newblock Probability Theory and Related Fields \textbf{97}, 543--558
\newblock  (1993)

\bibitem{Weiss(08)}
Wei\ss, C.: Thinning operations for modeling time series of counts---a survey.
\newblock AStA Advances in Statistical Analysis \textbf{92}, 319--341
\newblock  (2008)

\bibitem{WolpertBrown(11)}
Wolpert, R.L., Brown, L.D.: Stationary infinitely-divisible \uppercase{M}arkov
  processes with non-negative integer values
\newblock  (2011).
\newblock Working paper, Department of Staistics, Duke University

\bibitem{WolpertTaqqu(05)}
Wolpert, R.L., Taqqu, M.S.: Fractional {Ornstein-Uhlenbeck}
  \uppercase{L}\'{e}vy processes and the telecom process: Upstairs and
  downstairs.
\newblock Signal Processing \textbf{85}, 1523--1545
\newblock  (2005)

\bibitem{ZhuJoe(03)}
Zhu, R., Joe, H.: A new type of discrete self-decomposability and its
  application to continuous-time {Markov} processes for modeling count data
  time series.
\newblock Stochastic Models \textbf{19}, 235--254
\newblock  (2003)

\end{thebibliography}
